\documentclass[journal]{IEEEtran}
\usepackage{amssymb}
\usepackage[cmex10]{amsmath}
\usepackage{stfloats}
\usepackage{graphicx}
\usepackage{subfigure}
\usepackage{tabularx}
\usepackage{epsfig,epsf,color,balance,cite}
\usepackage{verbatim}
\usepackage{url}
\usepackage{bm}

\newtheorem{theorem}{Theorem}
\newtheorem{lemma}{Lemma}
\newtheorem{coro}{Corollary}
\newtheorem{remark}{Remark}
\usepackage{algorithm}
\usepackage{algpseudocode}
\begin{document}

\title{\textcolor[rgb]{0.00,0.00,0.00}{Pilot Reuse Among D2D Users in D2D Underlaid Massive MIMO Systems}}

\author{
\IEEEauthorblockN{Hao Xu, \emph{Student Member, IEEE}\IEEEauthorrefmark{0},
                  Wei Xu, \emph{Senior Member, IEEE}\IEEEauthorrefmark{0},
                  Zhaohui Yang, \emph{Student Member, IEEE}\IEEEauthorrefmark{0},}
                  {Jianfeng Shi, \emph{Student Member, IEEE}\IEEEauthorrefmark{0},
 and
                  Ming Chen, \emph{Member, IEEE}\IEEEauthorrefmark{0}
                  }
\thanks{Copyright (c) 2015 IEEE. Personal use of this material is permitted. However, permission to use this material for any other purposes must be obtained from the IEEE by sending a request to pubs-permissions@ieee.org. This work was in part supported by the NSFC (Nos. 61372106, 61471114, \& 61221002), NSTMP under 2016ZX03001016-003, the Six Talent Peaks project in Jiangsu Province under GDZB-005, Science and Technology Project of Guangdong Province under Grant 2014B010119001, the Scholarship from the China Scholarship Council (No. 201606090039), Program Sponsored for Scientific Innovation Research of College Graduate in Jiangsu Province under Grant KYLX16\_0221, and the Scientific Research Foundation of Graduate School of Southeast University under Grant YBJJ1651. \emph{(Corresponding author: Hao Xu, Wei Xu.)}}
\thanks{H. Xu, W. Xu, Z. Yang, J. Shi and M. Chen are with the National Mobile Communications Research
Laboratory, Southeast University, Nanjing 210096, China (Email: \{xuhao2013, wxu, yangzhaohui, shijianfeng and chenming\}@seu.edu.cn).}
}

\maketitle

\begin{abstract}
In a device-to-device (D2D) underlaid massive MIMO system, D2D transmitters reuse the uplink spectrum of cellular users (CUs), leading to cochannel interference. To decrease pilot overhead, we assume pilot reuse (PR) among  D2D pairs. We first derive the minimum-mean-square-error (MMSE) estimation of all channels and give a lower bound on the ergodic achievable rate of both cellular and D2D links. To mitigate pilot contamination caused by PR, we then propose a pilot scheduling and pilot power control algorithm based on the criterion of minimizing the sum mean-square-error (MSE) of channel estimation of D2D links. We show that, with an appropriate PR ratio and a well designed pilot scheduling scheme, each D2D transmitter could transmit its pilot with maximum power. In addition, we also maximize the sum rate of all D2D links while guaranteeing the quality of service (QoS) of CUs, and develop an iterative algorithm to obtain a suboptimal solution. Simulation results show that the effect of pilot contamination can be greatly decreased by the proposed pilot scheduling algorithm, and the PR scheme provides significant performance gains over the conventional orthogonal training scheme in terms of system spectral efficiency.
\end{abstract}

\IEEEpeerreviewmaketitle

\section{Introduction}
\label{section1}
With the increasing demand on broadband wireless communications, the problem of spectrum insufficiency has become a major factor limiting the wireless system performance \cite{force2002report}. Massive multiple-input multiple-output (MIMO) transmission was proposed in \cite{marzetta2010noncooperative} and has triggered considerable research interest recently due to its great gains in spectral efficiency (SE) and energy efficiency (EE) \cite{rusek2013scaling, larsson2014massive, xie2016full}. Besides, device-to-device (D2D) communication has also been proven promising in enhancing the SE of the traditional cellular systems and has drawn great attention recently \cite{liu2015device,asadi2014survey,lin2014overview}. Different from the conventional cellular communication where all traffic is routed via base station (BS), D2D communication allows two closely located users to communicate directly, and thus have distinct advantages such as high SE, short packet delay, low energy consumption and increased safety.

There has been extensive research on design and analysis of massive MIMO systems \cite{ngo2013energy, yang2013performance, hoydis2013massive, liang2015uplink, bai2013asymptotic}. In \cite{ngo2013energy}, the uplink capacity bounds were derived under both perfect and imperfect channel state information (CSI), and the tradeoff between SE and EE was studied. Ref. \cite{yang2013performance} compared two most prominent linear precoders with respect to (w.r.t.) SE and radiated EE in a massive MIMO system. Unlike \cite{ngo2013energy} and \cite{yang2013performance}, which considered simplified single-cell scenarios, \cite{hoydis2013massive, liang2015uplink, bai2013asymptotic} studied multi-cell massive MIMO systems. As for underlaid D2D communication, a great challenge to the existing cellular architecture is the cochannel interference due to spectrum reuse. There has been a lot of literature working on interference mitigation for D2D underlaid systems \cite{zhu2014downlink ,maghsudi2015joint, feng2013device, zhao2015resource, jiang2016energy,wang2015energy,hoang2015energy}. In \cite{zhu2014downlink ,maghsudi2015joint, feng2013device, zhao2015resource}, resource allocation and power control algorithms were proposed to maximize the SE of D2D users (DUs), and in \cite{jiang2016energy,wang2015energy,hoang2015energy}, extended algorithms were carried out to maximize the EE of DUs.

Though massive MIMO and D2D communication have been widely studied, only a few papers investigated the interplay between massive MIMO and D2D communication \cite{lin2015spectral, lin2014interplay}. In \cite{lin2014interplay}, the SE of cellular and D2D links was investigated under both perfect and imperfect CSI, but the overhead for acquiring CSI was not considered. In massive MIMO systems, orthogonal pilots are transmitted by cellular users (CUs) to obtain CSI. When D2D communication is introduced and orthogonal pilots are used at each D2D transmitter (D2D-Tx) for channel estimation, the pilot overhead is large which will significantly affect the system performance. Furthermore, as a multi-user transmission strategy, massive MIMO is designed to support multiple users transmitting on the same time-frequency block. Though D2D-to-cellular interference can be greatly reduced by a large antenna array at BS, cellular-to-D2D interference still persists and may be worse than a conventional single-input single-output (SISO) D2D underlaid system.

In order to shorten pilot overhead, an effective strategy is allowing orthogonal pilots to be reused among different users. Most of the existing works with pilot reuse (PR) mainly focus on multiple-cell scenarios, i.e., mobile users in the same cell use orthogonal pilots, and users in different cells reuse the same set of pilots \cite{jose2011pilot, yin2013coordinated, ashikhmin2012pilot, fernandes2013inter, ngo2012evd}. To the best of the authors' knowledge, only a few works have considered the strategy of PR within a cell \cite{you2015pilot, liu2015pilot, liu2016pilot, xu2017pilot}. In \cite{you2015pilot}, the authors analyzed the feasibility of PR over spatially correlated massive MIMO channels with constrained channel angular spreads. Authors of \cite{liu2015pilot} and \cite{liu2016pilot} allowed D2D-Txs to reuse the pilots of CUs and proposed an interference-aided minimum-mean-square-error (MMSE) detector to suppress the D2D-to-cellular interference. \cite{xu2017pilot} also studied a D2D underlaid massive MIMO system with PR, but the performance of CUs was left out of consideration for simplicity. In contrast to these existing works, our work analyzes the achievable rate of both cellular and D2D links under PR, and proposes pilot scheduling as well as power control algorithms to optimize the system performance. The main contributions of this paper are summarized as follows:

$\bullet$ We assume that CUs use orthogonal pilots while all D2D-Txs reuse another set of pilots for channel estimation. The motivation of PR stems from that D2D pairs usually locate dispersively and use low power for short-distance transmission. Hence, letting several D2D pairs which are far away from each other use the same pilot for channel estimation would cause endurable pilot contamination. Under PR, we first derive the expression of MMSE estimate of all channels. With the obtained imperfect CSI, all receivers apply the partial zero forcing (PZF) receive filters studied in \cite{jindal2011multi} for signal detection. Then, we derive the effective signal-to-interference-plus-noise ratio (SINR) and a lower bound on the ergodic achievable rate of each user.

$\bullet$ Different from the estimation of cellular channel vectors which is only affected by noise, the channel estimation of D2D links experiences effect from both noise and pilot contamination due to PR. To mitigate pilot contamination, we develop a pilot scheduling and pilot power control algorithm under the criterion of minimizing the sum mean-square-error (MSE) of channel estimation of D2D links. We first show that with an appropriate number of orthogonal pilots available for DUs and a well designed pilot scheduling scheme, each D2D-Tx should transmit its pilot using the maximum power. Then, we develop a heuristic pilot scheduling scheme to allocate pilots to DUs, and show that the sum MSE of channel estimation of D2D links can be decreased significantly.

$\bullet$ We study the sum SE maximization for D2D links while guaranteeing the quality of service (QoS) of CUs. Efficient power control algorithms often play an important role in reducing cochannel interference and reaping the potential benefits of D2D communication. However, these algorithms are usually carried out based on the knowledge of instantaneous CSI of all links \cite{zhu2014downlink ,maghsudi2015joint, feng2013device, zhao2015resource}. Apart from the computational complexity, such algorithms require BS to gather instantaneous CSI of all links, which is difficult for implementation. Therefore, in this paper, we consider performing the power control algorithm periodically at a coarser frame level granularity based on large-scale fading coefficients which vary slowly. Simulation results show that the proposed algorithm converges rapidly and obtains a much higher sum SE of D2D links compared to the typical orthogonal training scheme.

Note that since PR among D2D pairs in a D2D underlaid massive MIMO system has been less well researched, we consider a simplified single-input multiple-output (SIMO) transmission for D2D communication as \cite{lin2014interplay}, and mainly focus on analyzing the effect of PR on the system performance. As for MIMO transmission, similar results can be obtained by simply modifying the analysis and optimization of this manuscript. On the other hand, our analysis in this paper focuses on a single-cell scenario for the sake of clarity. Regarding the multi-cell massive MIMO system, there has been a lot of literature working on mitigating pilot contamination \cite{zhu2015graph, li2013spatial, mochaourab2016adaptive}. As a result, for a multi-cell D2D underlaid massive MIMO system, where PR among D2D pairs persists, we can first use the algorithms developed in \cite{zhu2015graph, li2013spatial, mochaourab2016adaptive} to allocate pilots to CUs if CSI can be exchanged among cells, and then straightforwardly extend the proposed algorithms to improve system performance.

In this paper, we follow the common notations. $\mathbb N$, $\mathbb R$ and $\mathbb C$ denote the set of natural numbers, the real space and the complex space, respectively. The boldface upper (lower) case letters are used to denote matrices (vectors). ${\bm I}_M$ stands for the $M \times M$ dimensional identity matrix and $\bm 0$ denotes the all-zero vector or matrix. `` $\setminus$ " represents the set subtraction operation. Superscript $(\cdot)^H$ denotes the conjugated-transpose operation and ${\mathbb E}\{\cdot\}$ denotes the expectation operation. We use $\left\| {\bm a} \right\|_2$ to denote the Euclidean norm of $\bm a$. ${\bm a}\succeq \bm 0 ({\bm a}\succ \bm 0)$ means that each element in ${\bm a}$ is positive (nonnegative).

The rest of this paper is organized as follows. In Section II, a D2D underlaid massive MIMO system is introduced. In Section III, we present the MMSE estimate of all channels under PR and show how PR affects the channel estimation. The achievable rate of both cellular and D2D links is analyzed in Section IV. In Section V, we aim to minimize the sum MSE of channel estimation of D2D links and maximize the sum SE of all D2D links. Finally, numerical verifications are presented in Section VI before concluding remarks in Section VII.

\section{System Model}
\label{section2}
Consider the uplink of a D2D underlaid massive MIMO system with one BS, $N$ CUs and $K$ D2D pairs. The set of CUs and D2D pairs are denoted by ${\cal N} =\{1,\cdots,N\}$\footnote{Here we misuse the notation ${\cal N}$ while avoiding possible ambiguity with the ${\cal N}$ in the complex normal distribution sign ${\cal CN}$.} and ${\cal{K}} =\{1,\cdots,K\}$, respectively. The BS is equipped with $B$ antennas and each CU has one transmit antenna. As for the D2D communication, we assume SIMO transmission, i.e., each D2D-Tx is equipped with one antenna and each D2D receiver (D2D-Rx) is equipped with $M$ antennas as in \cite{lin2014interplay}. In this system, all transmitters use the same time-frequency resource block to transmit signals, leading to cochannel interference. The $B\times 1$ dimensional received data vector at BS is
\begin{equation}
{\bm y}^{(\text{c})} \!=\! \sum\limits_{n = 1}^N \sqrt{q_{{\text s},n} u_{n}^{(\text{c})}} {\bm h}_{n}^{(\text{c})} x_n^{(\text{c})} \!+\! \sum\limits_{i = 1}^K \sqrt{p_{{\text s},i} u_{i}^{(\text{d})}} {\bm h}_{i}^{(\text{d})} x_i^{(\text{d})} \!+ {\bm z},
\label{received_data_BS}
\end{equation}
where $q_{{\text s},n}$ is the data transmit power of CU $n$, and $x_n^{(\text{c})}$ is the zero-mean unit-variance data symbol of CU $n$. $u_n^{(\text{c})}$ is the real-valued large-scale fading coefficient from CU $n$ to BS and is assumed to be known as a priori. ${\bm h}_n^{(\text{c})}\sim {\cal CN}({\bm 0},{\bm I}_B)$ denotes the fast fading vector channel from CU $n$ to BS. $p_{{\text s},i}$, $x_i^{(\text{d})}$, $u_{i}^{(\text{d})}$ and ${\bm h}_{i}^{(\text{d})}$ are similarly defined for D2D-Tx $i$. ${\bm z} \in {\mathbb C}^{B}$ is the complex Gaussian noise at BS with covariance $N_0 {\bm I}_B$.

Analogously, the $M\times 1$ dimensional received data vector at D2D-Rx $k$ is given by
\begin{equation}
{\bm y}_k^{(\text{d})} \!\!=\! \sum\limits_{i = 1}^K \!\sqrt{p_{{\text s},i} v_{ik}^{(\text{d})}} {\bm g}_{ik}^{(\text{d})} x_i^{(\text{d})} \!+\! \sum\limits_{n = 1}^N \!\sqrt{q_{{\text s},n} v_{nk}^{(\text{c})}} {\bm g}_{nk}^{(\text{c})} x_n^{(\text{c})}\!+ {\bm n}_k,
\label{received_data}
\end{equation}
where $v_{ik}^{(\text{d})}$ and ${\bm g}_{ik}^{(\text{d})}\sim {\cal CN}({\bm 0},{\bm I}_M)$ denote the real-valued large-scale fading coefficient and the fast fading vector channel from D2D-Tx $i$ to D2D-Rx $k$, respectively. $v_{nk}^{(\text{c})}$ and ${\bm g}_{nk}^{(\text{c})}$ are similarly defined for the link from CU $n$ to D2D-Rx $k$. ${\bm n}_k \in {\mathbb C}^{M}$ is the complex Gaussian noise at D2D-Rx $k$ with covariance $N_0 {\bm I}_M $.

\section{Channel Estimation}
\label{section3}
Orthogonal pilots are usually adopted to obtain the CSI of all links. In a D2D underlaid system, to reduce pilot overhead, we assume that CUs use orthogonal pilots while all D2D-Txs reuse another set of pilots for channel estimation. Denote ${\bm \Omega} = \left({\bm \omega}_1,\cdots,{\bm \omega}_N,{\bm \omega}_{N+1},\cdots,{\bm \omega}_{\tau}\right) \in $ ${\mathbb C}^{\tau \times \tau}$ as the pilot matrix with orthogonal column vectors (i.e., ${\bm \Omega}^H{\bm \Omega} \!=\! {\bm I}_{\tau}$). $\tau$ ($N < \tau \leq N+K$) is the length of the pilots and is also the number of pilots available for channel estimation (this is the smallest amount of pilots that are required). Then, without loss of generality, we assume that pilot ${\bm \omega}_n$ ($1\leq n \leq N$) is allocated to CU $n$ and the remaining pilots $\{{\bm \omega}_{N+1},\cdots,{\bm \omega}_{\tau}\}$ are reused among all D2D pairs.
\subsection{Channel Estimation at BS}
\label{III_A}
Similar as the uplink data transmission in (\ref{received_data_BS}), the $B\times \tau$ dimensional received signal matrix of pilot transmission at BS is
\begin{equation}
{\bm Y}^{(\text{c})} \!=\! \sum\limits_{n = 1}^N \!\sqrt{q_{{\text p},n} u_{n}^{(\text{c})}} {\bm h}_{n}^{(\text{c})} {\bm \omega}_n^H \!+\! \sum\limits_{i = 1}^K \!\sqrt{p_{{\text p},i} u_{i}^{(\text{d})}} {\bm h}_{i}^{(\text{d})} {\bm \lambda}_i^H \!+\! {\bm Z},
\label{received_pilot_BS}
\end{equation}
where $q_{{\text p},n}$ and $p_{{\text p},i}$ denote the pilot transmit powers of CU $n$ and D2D-Tx $i$, respectively. ${\bm \lambda}_i \in \{{\bm \omega}_{N+1},\cdots,{\bm \omega}_{\tau}\}$ is the pilot allocated to D2D pair $i$. ${\bm Z}$ is the noise matrix which consists of independently and identically distributed (i.i.d.) Gaussian elements with zero mean and variance $N_0$.
Then, the MMSE estimate of ${\bm h}_n^{(\text{c})}$ is given by \cite{kailath2000linear}
\begin{equation}
{\hat {\bm h}_n^{(\text{c})}} = \frac{\sqrt{q_{{\text p},n} u_n^{(\text{c})}}}{q_{{\text p},n} u_n^{(\text{c})}+N_0}{\bm Y}^{(\text{c})}{\bm \omega}_n, \forall n \in {\cal N}.
\label{estimation_CU}
\end{equation}
Given the channel estimate vector ${\hat {\bm h}_n^{(\text{c})}}$, we can express the true channel vector ${\bm h}_n^{(\text{c})}$ as ${\bm h}_n^{(\text{c})} = {\hat {\bm h}_n^{(\text{c})}} + {\tilde {\bm h}_n^{(\text{c})}}$, where the error vector ${\tilde {\bm h}_n^{(\text{c})}}$ represents the CSI uncertainty. Due to the property of MMSE estimation \cite{kailath2000linear}, ${\hat {\bm h}_n^{(\text{c})}}$ is statistically independent of ${\tilde {\bm h}_n^{(\text{c})}}$, and they follow
\begin{equation}
{\hat {\bm h}_n^{(\text{c})}}  \sim  {\cal CN}({\bm 0},\delta_n^{(\text{c})}{\bm I}_B), {\tilde {\bm h}_n^{(\text{c})}}  \sim  {\cal CN}({\bm 0},\varepsilon_n^{(\text{c})}{\bm I}_B),
\label{channel_esti_error1}
\end{equation}
where
\begin{equation}
\delta_n^{(\text{c})} = \frac{q_{{\text p},n} u_n^{(\text{c})}}{ q_{{\text p},n} u_n^{(\text{c})}+N_0 }, \varepsilon_n^{(\text{c})} = 1- \delta_n^{(\text{c})}.
\label{esti_var1}
\end{equation}

We then analogously derive the MMSE estimate of ${\bm h}_{k}^{(\text{d})}$ as follows
\begin{equation}
{\hat {\bm h}_{k}^{(\text{d})}} = \frac{\sqrt{p_{{\text p},k} u_{k}^{(\text{d})}}}{\sum\limits_{i \in {\cal X}_k} p_{{\text p},i} u_{i}^{(\text{d})}+ N_0} {\bm Y}^{(\text{c})}{\bm \lambda}_k, \forall k \in {\cal K},
\label{estimation_DU2BS}
\end{equation}
where ${\cal X}_k$ is the set of all D2D pairs using the same pilot as D2D pair $k$. Denote ${\bm h}_{k}^{(\text{d})}= {\hat {\bm h}_{k}^{(\text{d})}}+ {\tilde {\bm h}_{k}^{(\text{d})}}$. Then, ${\hat {\bm h}_{k}^{(\text{d})}}$ and ${\tilde {\bm h}_{k}^{(\text{d})}}$ are statistically independent satisfying
\begin{equation}
{\hat {\bm h}_{k}^{(\text{d})}}  \sim  {\cal CN}({\bm 0},\delta_{k}^{(\text{d})}{\bm I}_B), {\tilde {\bm h}_{k}^{(\text{d})}}  \sim  {\cal CN}({\bm 0},\varepsilon_{k}^{(\text{d})}{\bm I}_B),
\label{channel_esti_error2}
\end{equation}
where
\begin{equation}
\delta_{k}^{(\text{d})} = \frac{p_{{\text p},k} u_{k}^{(\text{d})}}{ \sum\limits_{i \in {\cal X}_k} p_{{\text p},i} u_{i}^{(\text{d})}+N_0 }, \varepsilon_{k}^{(\text{d})} = 1- \delta_{k}^{(\text{d})}.
\label{esti_var2}
\end{equation}
\subsection{Channel Estimation at D2D-Rxs}
The $M\times \tau$ dimensional received pilot signal matrix at D2D-Rx $k$ can be written as
\begin{equation}
{\bm Y}_k^{(\text{d})} \!=\! \sum\limits_{n = 1}^N \sqrt{q_{{\text p},n} v_{nk}^{(\text{c})}} {\bm g}_{nk}^{(\text{c})} {\bm \omega}_n^H \!+\! \sum\limits_{i = 1}^K \sqrt{p_{{\text p},i} v_{ik}^{(\text{d})}} {\bm g}_{ik}^{(\text{d})} {\bm \lambda}_i^H \!+\! {\bm N}_k,
\label{received_pilot}
\end{equation}
where ${\bm N}_{k}$ is the noise matrix consisting of i.i.d. Gaussian elements with zero mean and variance $N_0$. Then, the MMSE estimate of ${\bm g}_{ik}^{(\text{d})}$ is given by
\begin{equation}
{\hat {\bm g}_{ik}^{(\text{d})}} = \frac{\sqrt{p_{{\text p},i} v_{ik}^{(\text{d})}}}{{ \sum\limits_{j \in {\cal X}_k} p_{{\text p},j} v_{jk}^{(\text{d})}+N_0 }} {\bm Y}_k^{(\text{d})}{\bm \lambda}_i , \forall i,k \in {\cal K},
\label{channel_estimation}
\end{equation}
Denote ${\bm g}_{ik}^{(\text{d})} = {\hat {\bm g}_{ik}^{(\text{d})}} + {\tilde {\bm g}_{ik}^{(\text{d})}}$. Then, as mentioned above, ${\hat {\bm g}_{ik}^{(\text{d})}}$ is statistically independent of ${\tilde {\bm g}_{ik}^{(\text{d})}}$, and the distributions of them are given by
\begin{equation}
{\hat {\bm g}_{ik}^{(\text{d})}} \sim  {\cal CN}({\bm 0},\mu_{ik}^{(\text{d})}{\bm I}_M), {\tilde {\bm g}_{ik}^{(\text{d})}} \sim  {\cal CN}({\bm 0},\epsilon_{ik}^{(\text{d})}{\bm I}_M),
\label{channel_esti_error4}
\end{equation}
where
\begin{equation}
\mu_{ik}^{(\text{d})} = \frac{p_{{\text p},i} v_{ik}^{(\text{d})}}{ \sum\limits_{j \in {\cal X}_k} p_{{\text p},j} v_{jk}^{(\text{d})}+N_0 },\nonumber\\
\epsilon_{ik}^{(\text{d})} = 1- \mu_{ik}^{(\text{d})}.
\label{esti_var3}
\end{equation}

Similarly, we have the MMSE estimate of ${\bm g}_{nk}^{(\text{c})}$ as follows
\begin{equation}
{\hat {\bm g}_{nk}^{(\text{c})}} = \frac{\sqrt{q_{{\text p},n} v_{nk}^{(\text{c})}}}{q_{{\text p},n} v_{nk}^{(\text{c})}+ N_0} {\bm Y}_k^{(\text{d})}{\bm \omega}_n, \forall n \in {\cal N}, k \in {\cal K}.
\label{estimation_CU2DU}
\end{equation}
Denote  the channel estimation error vector by ${\tilde {\bm g}_{nk}^{(\text{c})}}$, then, ${\hat {\bm g}_{nk}^{(\text{c})}}$ and ${\tilde {\bm g}_{nk}^{(\text{c})}}$ are statistically independent and satisfy
\begin{equation}
{\hat {\bm g}_{nk}^{(\text{c})}} \sim  {\cal CN}({\bm 0},\mu_{nk}^{(\text{c})}{\bm I}_M), {\tilde {\bm g}_{nk}^{(\text{c})}} \sim  {\cal CN}({\bm 0},\epsilon_{nk}^{(\text{c})}{\bm I}_M),
\label{channel_esti_error}
\end{equation}
where
\begin{equation}
\mu_{nk}^{(\text{c})} = \frac{q_{{\text p},n} v_{nk}^{(\text{c})}}{q_{{\text p},n} v_{nk}^{(\text{c})}+N_0 }, \epsilon_{nk}^{(\text{c})} = 1- \mu_{nk}^{(\text{c})}.
\label{esti_var4}
\end{equation}

From (\ref{estimation_CU}) and (\ref{estimation_CU2DU}), it can be observed that the estimation of  ${\bm h}_n^{(\text{c})}$ and ${\bm g}_{nk}^{(\text{c})}$ is only affected by pilot noise. The pilot interference from other CUs and DUs disappear completely due to the orthogonality of the pilots. As for the estimation of ${\bm h}_k^{(\text{d})}$ and ${\bm g}_{ik}^{(\text{d})}$, it is clear from (\ref{estimation_DU2BS}) and (\ref{channel_estimation}) that apart from the effect of pilot noise, it is also affected by pilot contamination due to PR among D2D pairs.
\section{Achievable Rate Analysis}
\label{Section_IV}
In this section, we analyze the achievable rate of both cellular and D2D links under PR. Let $\bm \beta_n^{(\text{c})}$ denote the unit norm receive filter used by BS for detecting the signal of CU $n$, and $\bm \beta_k^{(\text{d})}$ denote the unit norm receive filter adopted by D2D-Rx $k$ for detecting the signal from D2D-Tx $k$. Since the receive filter can be used either to boost the desired signal power or to eliminate interference signal, the SINR of each link critically depends on the receive filter that is used. In this paper, we adopt PZF receivers, which use part of degrees of freedom for signal enhancement and the remaining degrees of freedom for interference suppression, for signal detection at both BS and D2D-Rxs.

We assume that BS uses $b_\text{c}$ and $b_\text{d}$ degrees of freedom to cancel the interference from the nearest $b_\text{c}$ cellular interferers and the nearest $b_\text{d}$ D2D interferers using different orthogonal pilots. From (\ref{estimation_DU2BS}), we can obtain the following relationship
\begin{equation}
{\hat {\bm h}_{k}^{(\text{d})}} = \sqrt {\frac{p_{{\text p},k} u_k^{(\text{d})}}{p_{{\text p},i} u_i^{(\text{d})}}} {\hat {\bm h}_i^{(\text{d})}}, \forall k \in {\cal K}, i \in {\cal X}_k \setminus k,
\end{equation}
which indicates that the estimation of channels from two different D2D-Txs using the same pilot to BS are in the same direction. As a result, the interference from D2D interferers applying the same pilot can be eliminated simultaneously by using one degree of freedom. Since $\tau-N$ orthogonal pilots are reused by $K$ D2D pairs, we divide all D2D pairs into $\tau-N$ sets with D2D pairs in each set using the same pilot for channel estimation. Then, we know that BS is able to cancel the interference from $b_\text{d}$ sets of D2D interferers. The feasible set of $(b_\text{c}, b_\text{d})$ is given by
\begin{equation}
\left\{(b_\text{c}, b_\text{d})\!\in\! {\mathbb N}\!\times\! {\mathbb N}\mid b_\text{c} \!\leq\! N\!-\!1, b_\text{d} \!\leq\! \tau\!-\!N, b_\text{c}\!+\!b_\text{d}\!\leq\! B\!-\!1 \right\}.
\label{feasible_b}
\end{equation}
The PZF filter $\bm \beta_n^{(\text{c})}$ can be obtained by normalizing the projection of channel estimation ${\hat {\bm h}_{n}^{(\text{c})}}$ on the nullspace of channel estimation vectors of cancelled interferers (refer to (\ref{PZF_CU}) in Appendix A). For the sake of convenience, let ${\cal C}_n^{(\text{c})}$ denote the set of uncancelled CUs when detecting $x_n^{(\text{c})}$, and ${\cal D}^{(\text{c})}$ denote the set of uncancelled DUs when detecting cellular signals.

Similarly, each D2D-Rx uses $m_\text{c}$ and $m_\text{d}$ degrees of freedom to cancel the interference from the nearest $m_\text{c}$ cellular interferers and the nearest $m_\text{d}$ D2D interferers using different orthogonal pilots. $(m_\text{c}, m_\text{d})$ should be in the following set
\begin{equation}
\left\{\!(\!m_\text{c}, m_\text{d})\!\in\! {\mathbb N}\!\times\! {\mathbb N}\!\mid\! m_\text{c} \!\leq\! N, m_\text{d} \!\leq\! \tau\!-\!N\!-\!1, m_\text{c}\!+\!m_\text{d}\!\leq\! M\!\!-\!\!1 \!\right\}\!.
\label{feasible_m}
\end{equation}
The PZF filter $\bm \beta_k^{(\text{d})}$ can be obtained by first projecting channel estimation ${\hat {\bm g}_{kk}^{(\text{d})}}$ onto the nullspace of channel estimation vectors of cancelled interferers, and then normalizing the projection. Let ${\cal C}_k^{(\text{d})}$ and ${\cal D}_k^{(\text{d})}$ respectively denote the sets of uncancelled CU and DUs when detecting $x_k^{(\text{d})}$.

Compared with MMSE receivers which optimally balance signal enhancement and interference suppression, PZF receivers are suboptimal. However, we adopt PZF receivers in this paper due to the following advantages. First, PZF receivers take both signal boosting and interference cancellation into account, which is similar as MMSE receivers in concept. It has been shown that the performance of PZF receivers in terms of system throughput can approach that of MMSE receivers and is much better than that of maximum ratio combining (MRC) or fully zero forcing (ZF) receivers \cite{jindal2011multi}. Second, the simple structure of PZF receivers makes the analysis of the system more tractable, and allows us to evaluate the performance of the D2D underlaid massive MIMO system in a more explicit way. In addition, PZF receivers can be simplified as MRC receivers by letting $(b_\text{c}, b_\text{d})=(0,0)$ or $(m_\text{c}, m_\text{d})=(0, 0)$, and can also be reduced to fully ZF receivers by letting $(b_\text{c}, b_\text{d})=(N-1, \tau-N)$ or $(m_\text{c}, m_\text{d})=(N, \tau-N-1)$, which can make the analysis of this paper more general. In order to cancel the nearest interferers and obtain PZF receivers, BS and D2D-Rxs have to know the positions of all transmitters. To relax this requirement and make it more practical, we can obtain PZF receivers by cancelling the interferers with the largest large-scale fading coefficients.
\subsection{A lower bound on achievable rate of cellular links}
Since BS only has the information of estimated channel vectors (\ref{estimation_CU}) and (\ref{estimation_DU2BS}), which are treated as the true CSI, using PZF receiver $\bm \beta_n^{(\text{c})}$ for detecting $x_n^{(\text{c})}$, we can write the post-processing received signal associated with CU $n$ at BS as
{\setlength\arraycolsep{2pt}
\begin{eqnarray}
\!\!\!\!\!\!&&r_n^{(\text{c})} =  \left({\bm \beta}_n^{(\text{c})}\right)^H {\bm y}^{(\text{c})}\nonumber\\
\!\!\!\!\!\!&&=  \sqrt{q_{{\text s},n} u_n^{(\text{c})}} \left({\bm \beta}_n^{(\text{c})}\right)^H \!{\hat {\bm h}_n^{(\text{c})}} x_n^{(\text{c})}\nonumber\\
\!\!\!\!\!\!&& + \left(\!{\bm \beta}_n^{(\text{c})}\!\right)^H \!\! \left(\sum\limits_{a \in {\cal C}_n^{(\text{c})}\setminus n} \!\!\!\sqrt{q_{{\text s},a} u_{a}^{(\text{c})}} {\hat {\bm h}_{a}^{(\text{c})}} x_a^{(\text{c})}\right. \!+\!\! \sum\limits_{i \in {\cal D}^{(\text{c})}}\!\! \sqrt{p_{{\text s},i} u_{i}^{(\text{d})}} {\hat {\bm h}_{i}^{(\text{d})}} x_i^{(\text{d})}\nonumber\\
\!\!\!\!\!\!&&  + \sum\limits_{a=1}^N\sqrt{q_{{\text s},a} u_{a}^{(\text{c})}} {\tilde {\bm h}_{a}^{(\text{c})}} x_a^{(\text{c})} +\left. \sum\limits_{i=1}^K \sqrt{p_{{\text s},i} u_{i}^{(\text{d})}} {\tilde {\bm h}_{i}^{(\text{d})}} x_i^{(\text{d})} + {\bm z}\!\right)\!,
\label{post_processing_signal_BS}
\end{eqnarray}}
\!\!\!where only the first term of the second equality is the desired information, while the other terms represent the cochannel interference, channel estimation error and noise, respectively. Thus, the effective SINR of cellular link $n$ can be expressed as
\begin{equation}
\eta_n^{(\text{c})} = \frac{S_n^{(\text{c})}}{ I_n^{(\text{c}\rightarrow \text{c})} + I_n^{(\text{d}\rightarrow \text{c})} + \alpha^{(\text{c})}\left\|{\bm \beta}_n^{(\text{c})}\right\|_2^2},
\label{SINR_CU}
\end{equation}
where $S_n^{(\text{c})}=q_{{\text s},n} u_n^{(\text{c})} \left|\left({\bm \beta}_n^{(\text{c})}\right)^H {\hat {\bm h}}_n^{(\text{c})}\right|^2$ represents the desired signal from CU $n$, $I_n^{(\text{c}\rightarrow \text{c})}$ and $I_n^{(\text{d}\rightarrow \text{c})}$ respectively denote the cochannel interference from all uncancelled cellular and D2D interferers, and they are given by
{\setlength\arraycolsep{2pt}
\begin{eqnarray}
&& I_n^{(\text{c}\rightarrow \text{c})}= \sum\limits_{a \in {\cal C}_n^{(\text{c})}\setminus n} q_{{\text s},a} u_a^{(\text{c})} \left|\left({\bm \beta}_n^{(\text{c})}\right)^H {\hat {\bm h}_a^{(\text{c})}}\right|^2,\nonumber\\
&& I_n^{(\text{d}\rightarrow \text{c})}=\sum\limits_{i\in {\cal D}^{(\text{c})}} p_{{\text s},i} u_i^{(\text{d})}  \left|\left({\bm \beta}_n^{(\text{c})}\right)^H {\hat {\bm h}_i^{(\text{d})}}\right|^2.
\label{abbre_CU}
\end{eqnarray}}
\!\!$\alpha^{(\text{c})}$ characterizes the effect of both channel estimation error and noise experienced by CU $n$, and can be formulated as
\begin{equation}
\alpha^{(\text{c})} =  \sum\limits_{a=1}^N q_{{\text s},a} u_a^{(\text{c})} \varepsilon_a^{(\text{c})} + \sum\limits_{i =1}^K p_{{\text s},i} u_i^{(\text{d})} \varepsilon_i^{(\text{d})} + N_0.
\end{equation}

In \cite{ngo2013energy} and \cite{lin2014interplay}, the asymptotic uplink rate of cellular links in a massive MIMO (or D2D underlaid massive MIMO) system has been studied. As for the system considered in this paper, we can also obtain a similar result about the asymptotic uplink rate of CUs as shown in the following corollary. Since corollary \ref{coro1} can be analogously verified as that in \cite{lin2014interplay}, we omit the proof process for brevity.
\begin{coro}
With fixed transmit powers at all transmitters, using linear filters for signal detection at BS, the asymptotic uplink rate of each cellular link grows unboundedly as $B$ goes to infinity.
\label{coro1}
\end{coro}

In fact, the number of antennas at BS is usually finite due to multiple practical constraints. Hence, in the following of this paper, we consider a more practical scenario where BS is equipped with large but finite numbers of antennas. Consider the block fading model, where all channels remain unchanged over the coherence interval with length $T$. Then, based on (\ref{channel_esti_error1}), (\ref{channel_esti_error2}) and (\ref{SINR_CU}), we can derive a lower bound on the ergodic achievable rate of cellular links as shown in the following theorem.
\begin{theorem}
Given the SINR formula in (\ref{SINR_CU}), the ergodic achievable rate of cellular link $n$ is lower bounded by
\begin{equation}
R_n^{(\text{c},{\text {lb}})}=\left(1-\frac{\tau}{T}\right)\log_2 \left( 1+\eta_n^{(\text{c},{\text {lb}})}\right), \forall n  \in {\cal N},
\label{lower_bound_CU}
\end{equation}
where
\begin{equation}
\eta_n^{(\text{c},{\text {lb}})}= \frac{q_{{\text s},n} \phi_n^{(\text{c})}}{\sum\limits_{a=1}^N q_{{\text s},a} \varphi_{an}^{(\text{c})} +\sigma^{(\text{c})}},
\label{SINR_lower_bound_CU}
\end{equation}
and
{\setlength\arraycolsep{2pt}
\begin{eqnarray}
\phi_n^{(\text{c})} &=& (B- b_\text{c} - b_\text{d} -1) u_n^{(\text{c})} \delta_n^{(\text{c})},\nonumber\\
\varphi_{an}^{(\text{c})} &=& \left\{
\begin{array}{ll}
u_a^{(\text{c})},&  a\in {\cal C}_n^{(\text{c})}\setminus n\\
u_a^{(\text{c})} \varepsilon_a^{(\text{c})},&  a=n\, {\text {or}}\, a\in {\cal N}\setminus {\cal C}_n^{(\text{c})}
\end{array} \right.,\nonumber\\
\sigma^{(\text{c})} &=& \sum\limits_{i=1}^K p_{{\text s},i} \varphi_i^{(\text{d})} + N_0, \nonumber\\
\varphi_i^{(\text{d})} &=& \left\{
\begin{array}{ll}
u_i^{(\text{d})},&  i\in {\cal D}^{(\text{c})}\\
u_i^{(\text{d})} \varepsilon_i^{(\text{d})},&  i\in {\cal K}\setminus {\cal D}^{(\text{c})}
\end{array} \right..
\end{eqnarray}}
\label{theorem2}
\end{theorem}

\itshape \textbf{Proof:}  \upshape
See Appendix A.
\hfill $\Box$

\begin{remark}
When considering the effect of pilot length $\tau$ on $R_n^{(\text{c},{\text {lb}})}$, from (\ref{SINR_lower_bound_CU}), we can find that when $b_\text{d}=0$, $\eta_n^{(\text{c},{\text {lb}})}$ is affected by channel estimation error of cellular link $n$, i.e., $\varepsilon_n^{(\text{c})}$, and channel estimation errors of $b_\text{c}$ cancelled cellular links, i.e., $\varepsilon_a^{(\text{c})}, a\in {\cal N}\setminus {\cal C}_n^{(\text{c})}$. Since CUs use orthogonal pilots for channel estimation, the value of $\tau$ has no effect on $\varepsilon_n^{(\text{c})}$ and $\varepsilon_a^{(\text{c})}, a\in {\cal N}\setminus {\cal C}_n^{(\text{c})}$. Hence, for fixed pilot transmit power, $R_n^{(\text{c},{\text {lb}})}$ decreases monotonically w.r.t. $\tau$. In contrast, when $b_\text{d}>0$, except the effect of $\varepsilon_n^{(\text{c})}$ and $\varepsilon_a^{(\text{c})}, a\in {\cal N}\setminus {\cal C}_n^{(\text{c})}$, $\eta_n^{(\text{c},{\text {lb}})}$ is also influenced by the estimation errors of channels from $b_\text{d}$ cancelled D2D-Txs to BS, i.e., $\varepsilon_i^{(\text{d})},  i\in {\cal K}\setminus {\cal D}^{(\text{c})}$. Due to PR, increasing $\tau$ results in smaller $\varepsilon_i^{(\text{d})},  i\in {\cal K}\setminus {\cal D}^{(\text{c})}$, and thereby helps increase $\eta_n^{(\text{c},{\text {lb}})}$. Therefore, it would be hard to determine the monotonicity of $R_n^{(\text{c},{\text {lb}})}$ w.r.t. $\tau$.
\label{remark1}
\end{remark}
\subsection{A lower bound on achievable rate of D2D links}
To detect $x_k^{(\text{d})}$, the received signal at D2D-Rx $k$ after using PZF receiver ${\bm \beta}_k^{(\text{d})}$ can be written as
{\setlength\arraycolsep{2pt}
\begin{eqnarray}
\!\!\!\!\!\!&&r_k^{(\text{d})} = \left({\bm \beta}_k^{(\text{d})}\right)^H {\bm y}_k^{(\text{d})}\nonumber\\
\!\!\!\!\!\!&&= \sqrt{p_{{\text s},k} v_{kk}^{(\text{d})}} \left({\bm \beta}_k^{(\text{d})}\right)^H \!{\hat {\bm g}_{kk}^{(\text{d})}} x_k^{(\text{d})}\nonumber\\
\!\!\!\!\!\!&&+ \left(\!{\bm \beta}_k^{(\text{d})}\!\right)^H \!\! \left(\!\sum\limits_{i \in {\cal D}_k^{(\text{d})}\setminus k}\!\! \sqrt{p_{{\text s},i} v_{ik}^{(\text{d})}} {\hat {\bm g}_{ik}^{(\text{d})}} x_i^{(\text{d})} \right. \!+\! \sum\limits_{n \in {\cal C}_k^{(\text{d})}}\! \sqrt{q_{{\text s},n} v_{nk}^{(\text{c})}} {\hat {\bm g}_{nk}^{(\text{c})}} x_n^{(\text{c})}\nonumber\\
\!\!\!\!\!\!&& + \sum\limits_{i=1}^K \sqrt{p_{{\text s},i} v_{ik}^{(\text{d})}} {\tilde {\bm g}_{ik}^{(\text{d})}} x_i^{(\text{d})} +\left. \sum\limits_{n=1}^N\sqrt{q_{{\text s},n} v_{nk}^{(\text{c})}} {\tilde {\bm g}_{nk}^{(\text{c})}} x_n^{(\text{c})} + {\bm n}_k\!\right)\!\!.
\label{post_processing_signal}
\end{eqnarray}}
\!\!\!where only the first term of the second equality is the desired signal, while the other terms respectively denote the cochannel interference, channel estimation error and noise. Then, the effective SINR of D2D link $k$ is
\begin{equation}
\eta_k^{(\text{d})} \!=\! \frac{S_k^{(\text{d})}}{I_k^{(\text{d}\rightarrow \text{d})} + I_k^{(\text{c}\rightarrow \text{d})} + \alpha_k^{(\text{d})} \left\|{\bm \beta}_k^{(\text{d})}\right\|_2^2},
\label{SINR}
\end{equation}
where $S_k^{(\text{d})}=p_{{\text s},k} v_{kk}^{(\text{d})} \left| \left({\bm \beta}_k^{(\text{d})}\right)^H {\hat {\bm g}}_{kk}^{(\text{d})}\right|^2$ denotes the desired signal from D2D-Rx $k$. $I_k^{(\text{d}\rightarrow \text{d})}$, $I_k^{(\text{c}\rightarrow \text{d})}$ respectively denote D2D and cellular cochannel interference, $\alpha_k^{(\text{d})}$ characterizes the effect of both channel estimation error and noise experienced by D2D-Rx $k$, and they are given by
{\setlength\arraycolsep{2pt}
\begin{eqnarray}
&& I_k^{(\text{d}\rightarrow \text{d})}=\sum\limits_{i \in {\cal D}_k^{(\text{d})}\setminus k} p_{{\text s},i} v_{ik}^{(\text{d})} \left|\left({\bm \beta}_k^{(\text{d})}\right)^H {\hat {\bm g}_{ik}^{(\text{d})}}\right|^2, \nonumber\\
&& I_k^{(\text{c}\rightarrow \text{d})}=\sum\limits_{n\in {\cal C}_k^{(\text{d})}} q_{{\text s},n} v_{nk}^{(\text{c})} \left|\left({\bm \beta}_k^{(\text{d})}\right)^H {\hat {\bm g}_{nk}^{(\text{c})}}\right|^2,\nonumber\\
&& \alpha_k^{(\text{d})} = \sum\limits_{i=1}^K p_{{\text s},i} v_{ik}^{(\text{d})} \epsilon_{ik}^{(\text{d})} + \sum\limits_{n=1}^N q_{{\text s},n} v_{nk}^{(\text{c})} \epsilon_{nk}^{(\text{c})} + N_0.
\label{abbre_DU}
\end{eqnarray}}
\!\!Similarly as the cellular uplink case, we can also derive a lower bound on the ergodic achievable rate of D2D links.
\begin{theorem}
Given the SINR formula in (\ref{SINR}), the ergodic achievable rate of D2D link $k$ is lower bounded by
\begin{equation}
R_k^{(\text{d},{\text {lb}})}=\left(1-\frac{\tau}{T}\right) \log_2 \left( 1+\eta_k^{(\text{d},{\text {lb}})}\right), \forall k \in {\cal K},
\label{lower_bound}
\end{equation}
where
\begin{equation}
\eta_k^{(\text{d},{\text {lb}})}=\frac{p_{{\text s},k} \phi_k^{(\text{d})}}{\sum \limits_{i=1}^K p_{{\text s},i} \psi_{ik}^{(\text{d})} + \sigma_k^{(\text{d})}},
\label{SINR_lower_bound_DU}
\end{equation}
and
{\setlength\arraycolsep{2pt}
\begin{eqnarray}
\phi_k^{(\text{d})} &=& (M- m_\text{c} - m_\text{d}-1) v_{kk}^{(\text{d})} \mu_{kk}^{(\text{d})},\nonumber\\
\psi_{ik}^{(\text{d})} &=& \left\{
\begin{array}{ll}
v_{ik}^{(\text{d})}, \quad i\in {\cal D}_k^{(\text{d})}\setminus {\cal X}_k& \\
v_{ik}^{(\text{d})} \epsilon_{ik}^{(\text{d})}, \quad i=k \,{\text {or}}\, {\cal K}\setminus {\cal D}_k^{(\text{d})}&  \\
(M\!-\! m_\text{c} \!-\! m_\text{d}\!-\!1)v_{ik}^{(\text{d})} \mu_{ik}^{(\text{d})} \!+\! v_{ik}^{(\text{d})} \epsilon_{ik}^{(\text{d})},&  i\in {\cal X}_k\setminus k
\end{array} \right.,\nonumber\\
\sigma_k^{(\text{d})} &=& \sum\limits_{n \in {\cal C}_k^{(\text{d})}} q_{{\text s},n} v_{nk}^{(\text{c})} + \sum\limits_{n \in {\cal N} \setminus {\cal C}_k^{(\text{d})}} q_{{\text s},n} v_{nk}^{(\text{c})} \epsilon_{nk}^{(\text{c})} + N_0.
\label{simplify}
\end{eqnarray}}
\label{theorem3}
\end{theorem}

\itshape \textbf{Proof:}  \upshape
See Appendix B.
\hfill $\Box$

\begin{remark}
From (\ref{SINR_lower_bound_DU}), it can be found that for any $(m_\text{c}, m_\text{d})$ in feasible set (\ref{feasible_m}), $\eta_k^{(\text{d},{\text {lb}})}$ is an implicit function of $\tau$ due to PR. Increasing $\tau$ results in more accurate channel estimations of D2D links, and thereby helps increase $\eta_k^{(\text{d},{\text {lb}})}$. However, as $\tau$ increases, the number of symbols available for data transmission becomes smaller. In Section VI, we show by simulation results that $R_k^{(\text{d},{\text {lb}})}$ first increases and then decreases w.r.t. $\tau$.
\label{remark2}
\end{remark}

In the following, we focus on pilot scheduling and power control design based on (\ref{lower_bound_CU}) and (\ref{lower_bound}). Since large-scale fading coefficients vary slowly, the proposed pilot scheduling and power control algorithms can be performed periodically at a coarser frame level granularity, which will greatly decrease the computational complexity of BS. As a result, Theorem \ref{theorem2} and Theorem \ref{theorem3} are helpful for the following analysis.
\section{Pilot Scheduling and Power Control}
Up to now, we have investigated the cannel estimation as well as achievable rate of both cellular and D2D links in a D2D underlaid massive MIMO system with PR. Based on the above analysis, we focus on two problems in this section. The first problem aims to minimize the sum MSE of channel estimation of D2D links, and the second problem aims to maximize the sum rate of all D2D links while guaranteeing the QoS requirements of CUs.
\subsection{Pilot Power Control and Pilot Scheduling}
As mentioned in Section \ref{section3}, the channel estimation of cellular links is only affected by additive noise. Therefore, we assume that each CU transmits pilot signal with the maximum power. As for D2D links, apart from the effect of additive noise, channel estimation is also influenced by pilot contamination. Due to the location dispersion of D2D pairs and short-distance D2D transmission, it should be preferred that the pilot contamination can be greatly reduced by using an effective pilot scheduling and pilot power control algorithm. According to (\ref{channel_esti_error4}) and (\ref{esti_var3}), the sum MSE of channel estimation of D2D links can be written as
\begin{equation}
\sum \limits_{k=1}^K {\mathbb E} \left\{ \left\|{{\tilde {\bm g}_{kk}^{(\text{d})}}}\right\|_2^2\right\} = \sum \limits_{k=1}^K M \epsilon_{kk}^{(\text{d})}.
\label{sum_MMSE}
\end{equation}
Since orthogonal pilots $\{{\bm \omega}_{N+1},\cdots,{\bm \omega}_{\tau}\}$ are reused among $K$ D2D pairs, denote the PR pattern by ${\bm O}\in {\mathbb R}^{(\tau-N)\times K}$ with each element in ${\bm O}$ being binary-valued. If D2D pair $k$ is assigned pilot ${\bm \omega}_t \in \{{\bm \omega}_{N+1},\cdots,{\bm \omega}_{\tau}\}$, we have $o_{t-N,k}=1$, otherwise, we have $o_{t-N,k}=0$. Denote the pilot transmit power vector of DUs by ${\bm p}_{\text p} \triangleq \left[{p_{{\text p},1},\cdots,p_{{\text p},K}}\right]^T$. Then, aiming at minimizing (\ref{sum_MMSE}), we arrive at the following problem
{\setlength\arraycolsep{2pt}
\begin{subequations}
\begin{align}
\mathop {\min }\limits_{{\bm O},{\bm p}_{\text p}} \quad& \sum \limits_{k=1}^K M \epsilon_{kk}^{(\text{d})}  \label{MMSE_a}\\
\text{s.t.} \quad\; & 0 \leq p_{{\text p},k} \leq \tau P_k, \forall k \in {\cal K},\label{MMSE_b}\\
& \sum \limits_{t=N+1}^\tau o_{t-N,k} =1, \forall k \in {\cal K}, \label{MMSE_c}
\end{align}
\label{MMSE}
\end{subequations}}
\!\!\!\!\!where $P_k$ is the maximum data transmit power of D2D-Tx $k$. Constraints (\ref{MMSE_c}) indicate that each D2D pair can be allocated only one pilot. Note that for simplicity, $\epsilon_{kk}^{(\text{d})}$ is formulated as a function of ${\bm O}$ in an implicit way in (\ref{esti_var3}). We can also equivalently rewrite it in an explicit way as follows
\begin{equation}
\epsilon_{kk}^{(\text{d})} \!=\!\! \sum \limits_{t=N+1}^\tau \!\!o_{t-N,k}\! \left(\!1- \frac{p_{{\text p},k} v_{kk}^{(\text{d})}}{ \sum\limits_{j =1}^K o_{t-N,j} p_{{\text p},j} v_{jk}^{(\text{d})}\!+\!N_0 }\!\right), \forall k \in {\cal K}.
\end{equation}

To solve problem (\ref{MMSE}), we first give the optimal condition for ${\bm p}_{\text p}$ in the following theorem.
\begin{theorem}
There always exists $\tau \leq N+K$ such that when the optimal pilot scheduling matrix ${\bm O}^\text {opt}$ has been determined, the optimal ${\bm p}_{\text p}^\text {opt}$ satisfies $p_{{\text p},k}^\text {opt}=\tau P_k, \forall k \in {\cal K}$.
\label{theorem4}
\end{theorem}

\itshape \textbf{Proof:}  \upshape
See Appendix C.
\hfill $\Box$
The above theorem indicates that with a proper $\tau$ and the optimal ${\bm O}^\text {opt}$, constraints (\ref{MMSE_b}) are always active. We can explain Theorem $\ref{theorem4}$ in an intuitive way as follows. When the number of orthogonal pilots available for DUs is appropriate (i.e., with a relatively low PR ratio) and these pilots are allocated to D2D pairs by using the optimal pilot scheduling scheme (a special case is $\tau=N+K$ and all D2D pairs use different orthogonal pilots for channel estimation), ignorable pilot contamination would be caused due to the dispersive positions of D2D pairs. Hence, all D2D-Txs should transmit their pilot signals in the maximum power to increase the estimation accuracy. In contrast, with a small $\tau$ (i.e., with a relatively high PR ratio), even using the optimal pilot scheduling scheme, pilot contamination may still influence the estimation accuracy greatly and solving (\ref{MMSE7}) may yield $p_{{\text p},k}^\text {opt}=0$. In this case, we need to enlarge the set of pilots for DUs to decrease PR ratio.

Based on Theorem \ref{theorem4}, in the following, we assume that D2D-Txs always transmit pilots in the maximum power. Then, problem (\ref{MMSE}) becomes
{\setlength\arraycolsep{2pt}
\begin{subequations}
\begin{align}
\mathop {\min }\limits_{{\bm O}} \quad& \sum \limits_{k=1}^K M \epsilon_{kk}^{(\text{d})} \\
\text{s.t.} \quad\; & \sum \limits_{t=N+1}^\tau o_{t-N,k} =1, \forall k \in {\cal K},
\end{align}
\end{subequations}}
\!\!\!which is a mixed integer programming problem. The optimal ${\bm O}$ can be obtained through exhaustive search (ES). Recalling (\ref{esti_var3}), the number of scalar multiplication required to compute the objective function in (\ref{MMSE}) is ${\cal O}(K)$. Thus, obtaining the optimal ${\bm O}$ through ES involves a complexity of ${\cal O}(K (\tau-N)^K)$. Due to the exponential complexity, it will be impractical to run ES when the number of D2D pairs is large. Therefore, we propose a low complexity pilot scheduling algorithm.

To mitigate pilot contamination in a multi-cell massive MIMO system, \cite{zhu2015graph} proposed the GCPA scheme, in which a metric is defined to indicate the interference strength among CUs and a binary matrix is used to describe the connections of CUs. However, to obtain the binary matrix, a suboptimal threshold needs to be found by applying iterative grid search. In this paper, we define a continuous-valued metric $\chi_{ik}$ to evaluate the potential interference strength between D2D pair $i$ and $k$
\begin{equation}
\chi_{ik} = \left\{
\begin{array}{ll}
\!\!0,& \!\! \forall k \in {\cal K}, i= k\\
\!\!\ln \left(1\!+\!\left(\frac{v_{ik}^{(\text{d})}}{v_{kk}^{(\text{d})}}\right)^2\!+\!
\left(\frac{v_{ki}^{(\text{d})}}{v_{ii}^{(\text{d})}}\right)^2\right),& \!\! \forall k \in {\cal K}, i\in {\cal K}\setminus k\\
\end{array} \right.\!\!.
\label{inter_strength}
\end{equation}
Denote $\Lambda$ as the set of D2D pairs which have been allocated pilots, then, we summarize the pilot scheduling algorithm in Algorithm \ref{PSA}.
\begin{algorithm}[h]
\begin{algorithmic}
\caption{Pilot Scheduling Algorithm (PSA)}
\label{PSA}
\State \textbf {Initialization:}
\State \quad $\Lambda=\emptyset, {\bm O}={\bm 0} $. Calculate $\chi_{ik}, \forall i,k \in {\cal K}$.
\State \textbf {Pilot Allocation:}
\For {$j=1,\cdots, K$}
\vspace{0.3em}
\State 1: $k^{'}= \arg \mathop { \max }\limits_{k \in {\cal K}\setminus \Lambda} \sum\limits_{i \in {\cal K}} \chi_{ik}$,
\vspace{0.3em}
\State 2: $t^{'}=\arg \mathop { \min }\limits_{t \in \{N+1,\cdots,\tau\}} \sum\limits_{\left\{i|{\bm \lambda}_i={\bm \omega}_t\right\}} \chi_{ik^{'}}$,
\vspace{0.3em}
\State 3: ${\bm \lambda}_{k^{'}}= {\bm \omega}_{t^{'}}$, $o_{t^{'}-N,k^{'}} = 1$, $\Lambda=\Lambda\cup k^{'}$.
\vspace{0.3em}
\EndFor
\end{algorithmic}
\end{algorithm}

The basic idea of the PSA algorithm is that the D2D pair experiencing larger pilot contamination possesses a higher priority for pilot allocation. The main steps in each iteration can be explained as follows. First, D2D pair $k^{'} \in {\cal K}\setminus \Lambda$ experiencing the largest potential interference from other DUs is selected. Then, the pilot causing the least interference to $k^{'}$ is chosen. Finally, pilot ${\bm \omega}_{t^{'}}$ is assigned to D2D pair $k^{'}$, i.e., $o_{t^{'}-N,k^{'}} = 1$, and $\Lambda$ is updated by $\Lambda=\Lambda\cup k^{'}$. The algorithm will be carried out for $K$ times until all D2D pairs are allocated with pilots.

\subsection{Data Power Control}
\label{section_Power_Control}
In this subsection, we aim to maximize the sum rate of all DUs while guaranteeing the QoS requirements of CUs. Since the exact expressions of ${\mathbb E}\left\{ R_n^{(\text{c})}\right\}$ and ${\mathbb E}\left\{ R_k^{(\text{d})}\right\}$  are unapproachable, we use their lower bounds (\ref{lower_bound_CU}) and (\ref{lower_bound}) for replacement. Simulation results show that the gap between the ergodic achievable rate and its lower bound is marginal, verifying the feasibility of the approximation.

Denote the data transmit power vectors of CUs and DUs by ${\bm q}_{\text s} \triangleq \left[{q_{{\text s},1},\cdots,q_{{\text s},N}}\right]^T$ and ${\bm p}_{\text s} \triangleq \left[{p_{{\text s},1},\cdots,p_{{\text s},K}}\right]^T$, respectively. Then, we arrive at the following problem
{\setlength\arraycolsep{2pt}
\begin{subequations}
\begin{align}
\mathop {\max }\limits_{{\bm q}_{\text s}, {\bm p}_{\text s}} \quad& \sum \limits_{k=1}^K R_k^{(\text{d},{\text {lb}})}  \label{power_control_a}\\
\text{s.t.} \quad\; & \eta_n^{(\text{c},{\text {lb}})}\geq \gamma_n, \forall n \in {\cal N},\label{power_control_b}\\
& 0 \leq q_{{\text s},n} \leq Q_n, \forall n \in {\cal N}, \label{power_control_c}\\
& 0 \leq p_{{\text s},k} \leq P_k, \forall k \in {\cal K},\label{power_control_d}
\end{align}
\label{power_control}
\end{subequations}}
\!\!\!\!\!\!where $\gamma_n$ and $Q_n$ respectively denote the target SINR and the maximum data transmit power of CU $n$. From (\ref{SINR_lower_bound_CU}) and (\ref{lower_bound}), we can see that ${\bm q}_{\text s}$ and ${\bm p}_{\text s}$ are coupled in the expressions of $\eta_n^{(\text{c},{\text {lb}})}$ and $R_k^{(\text{d},{\text {lb}})}$. Moreover, the fractional structure of SINR expressions and the log($\cdot$) operation make (\ref{power_control}) non-convex. Therefore, it is difficult to obtain the optimal solution of (\ref{power_control}). In the following, we divide the optimization into two consecutive parts. In the first part, we optimize ${\bm q}_{\text s}$ with ${\bm p}_{\text s}$ fixed, and vice versa for the other part.
\subsubsection{Data Power Control for Cellular Links}
For given ${\bm p}_{\text s}$, problem (\ref{power_control}) can be rewritten as
{\setlength\arraycolsep{2pt}
\begin{subequations}
\begin{align}
\mathop {\max }\limits_{{\bm q}_{\text s}} \quad& \sum \limits_{k=1}^K R_k^{(\text{d},{\text {lb}})}  \label{power_control1_a}\\
\text{s.t.} \quad\; & \eta_n^{(\text{c},{\text {lb}})}\geq \gamma_n, \forall n \in {\cal N},\label{power_control1_b}\\
& 0 \leq q_{{\text s},n} \leq Q_n, \forall n \in {\cal N}.\label{power_control1_c}
\end{align}
\label{power_control1}
\end{subequations}}
\!\!\!Based on (\ref{SINR_lower_bound_CU}), we can write SINR constraints (\ref{power_control1_b}) in a vector form as
\begin{equation}
{\bm q}_{\text s} \succeq {\bm F}{\bm q}_{\text s} + {\bm \theta} \triangleq {\bm \Delta}({\bm q}_{\text s}),
\label{first_constraint}
\end{equation}
where
\begin{displaymath}
\mathbf{F} =
\left( \begin{array}{ccc}
\frac{\gamma_1 \varphi_{11}^{(\text{c})}}{\phi_{1}^{(\text{c})}} & \ldots & \frac{\gamma_1 \varphi_{N1}^{(\text{c})}}{\phi_{1}^{(\text{c})}} \\
\vdots & \ddots & \vdots \\
\frac{\gamma_N \varphi_{1N}^{(\text{c})}}{\phi_N^{(\text{c})}} & \ldots & \frac{\gamma_N \varphi_{NN}^{(\text{c})}}{\phi_N^{(\text{c})}}
\end{array} \right),
\end{displaymath}
{\setlength\arraycolsep{2pt}
\begin{eqnarray}
&&{\bm \theta}=\left[\frac{\gamma_1 \sigma^{(\text{c})}}{\phi_1^{(\text{c})}},\cdots,\frac{\gamma_N \sigma^{(\text{c})}}{\phi_N^{(\text{c})}} \right]^T,\nonumber\\
&&{\bm \Delta}({\bm q}_{\text s})=\left[\Delta_1({\bm q}_{\text s}), \cdots, \Delta_N({\bm q}_{\text s})\right]^T.
\label{mediate_para}
\end{eqnarray}}
\!\!In (\ref{mediate_para}), ${\bm \theta}$ consists of scaled cochannel interference from D2D interferers and additive noise, and ${\bm \Delta}({\bm q}_{\text s})$ can be seen as an interference function \cite{chen2007uplink}.

\begin{remark}
If problem (\ref{power_control1}) is feasible, the SINR constraint of each CU would hold with equality for the optimal ${\bm q}_{\text s}^{\text {opt}}$. Equivalently, for the vector form (\ref{first_constraint}), we have ${\bm q}_{\text s}^{\text {opt}} = {\bm F}{\bm q}_{\text s}^{\text {opt}} + {\bm \theta}$. This can be readily verified by reductio. Assume that $\eta_n^{(\text{c},{\text {lb}})} > \gamma_n$ for ${\bm q}_{\text s}^{\text {opt}}$, then, we can further increase the objective function by decreasing $q_{{\text s},n}^{\text {opt}}$ slightly.
\label{remark4}
\end{remark}

According to \cite{chen2007uplink}, if the spectral radius of ${\bm F}$ is less than 1, the optimal power vector has the form ${\bm q}_{\text s}^{\text {opt}} = \left({\bm I}_N - {\bm F}\right)^{-1} {\bm \theta}$ . To obtain ${\bm q}_{\text s}^{\text {opt}}$, matrix inversion and spectral radius calculation are required, resulting in high complexity. As a result, we obtain ${\bm q}_{\text s}^{\text {opt}}$ by using the following low complexity iterative scheme
\begin{equation}
{\bm q}_{\text s}(l+1)={\bm \Lambda} ({\bm q}_{\text s}(l)),
\label{iteration}
\end{equation}
where $l$ represents the time instant, and ${\bm \Lambda} ({\bm q}_{\text s}(l)) = \left\{\min \left\{ Q_1, \Delta_1 ({\bm q}_{\text s}(l))\right\}, \cdots, \min \left\{ Q_N, \Delta_N ({\bm q}_{\text s}(l))\right\}\right\}$. We denote this data power control algorithm for cellular links by DPCC. By proving that ${\bm \Lambda} ({\bm q}_{\text s})$ is standard \cite{yates1995framework}, we can readily verify that the DPCC algorithm converges to the optimal solution for any initial power vector ${\bm q}_{\text s} \succeq {\bm 0}$. The detailed proof is given in Appendix D.
\subsubsection{Data Power Control for D2D Links}
For fixed ${\bm q}_{\text s}$, we arrive at the following problem. For notational simplicity, the constant coefficient $1-\frac{\tau}{T}$ in (\ref{lower_bound}) is omitted.
{\setlength\arraycolsep{2pt}
\begin{subequations}
\begin{align}
\mathop {\max }\limits_{{\bm p}_{\text s}} \quad& \sum \limits_{k=1}^K \log_2 \left( 1+\frac{p_{{\text s},k} \phi_k^{(\text{d})}}{\sum \limits_{i=1}^K p_{{\text s},i} \psi_{ik}^{(\text{d})} + \sigma_k^{(\text{d})}}\right)  \label{power_control2_a}\\
\text{s.t.} \quad\; & \sum\limits_{k=1}^K p_{{\text s},k} \varphi_k^{(\text{d})} \leq \zeta,\label{power_control2_b}\\
& 0 \leq p_{{\text s},k} \leq P_k, \forall k \in {\cal K}, \label{power_control2_c}
\end{align}
\label{power_control2}
\end{subequations}}
\!\!\!\!\!\!\!where $\zeta=\min\left\{\zeta_1,\cdots,\zeta_N\right\}$, and $\zeta_n=\frac{1}{\gamma_n} q_{{\text s},n} \phi_n^{(\text{c})}-\sum\limits_{a=1}^N q_{{\text s},a} \varphi_{an}^{(\text{c})} -N_0$.

It can be directly seen that problem (\ref{power_control2}) is non-convex due to the fractional expression of $\eta_k^{(\text{d},{\text {lb}})}$ and the $\log(\cdot)$ operation in (\ref{power_control2_a}). In order to solve (\ref{power_control2}), we transform it into the equivalent form in (\ref{equivalent_problem2}) which admits suboptimal solutions using the well-known WMMSE approach in \cite{shi2011iteratively}. The details are provided in Appendix E. We summarize the data power control algorithm for D2D links (labeled as `DPCD') in Algorithm \ref{algorithm2}.
\begin{algorithm}[h]
\begin{algorithmic}[1]
\caption{ Data Power Control Algorithm for D2D Links (DPCD)}
\label{IRAPC}
\State Initialize $f_k=\sqrt{P_k}, w_k=1, \forall k \in {\cal K}$, and accuracy $\rho_1$.
\Repeat
\vspace{0.3em}
\State \!\!\!\!\!\!$w_k'=w_k, \forall k \in {\cal K},$
\vspace{0.3em}
\State \!\!\!\!\!\!$\nu_k = \frac{f_k \sqrt{\phi_k^{(\text{d})}}}{f_k^2 \phi_k^{(\text{d})}+\sum\limits_{i=1}^K f_i^2 \psi_{ik}^{(\text{d})} + \sigma_k^{(\text{d})}}, \forall k \in {\cal K},$
\vspace{0.3em}
\State \!\!\!\!\!\!$w_k = \left(1-\nu_k f_k \sqrt{\phi_k^{(\text{d})}}\right)^{-1}, \forall k \in {\cal K},$
\vspace{0.3em}
\State \!\!\!\!\!\!Obtain $\lambda$ by using the bisection method,
\vspace{0.3em}
\State \!\!\!\!\!\!$f_k\!=\! \min  \left\{\!\!{\sqrt {P_k}}, \frac{w_k \nu_k \sqrt{\phi_k^{(\text{d})}}}{w_k \nu_k^2 \phi_k^{(\text{d})} + \sum\limits_{i=1}^K w_i \nu_i^2 \psi_{ki}^{(\text{d})} + \lambda \varphi_k^{(\text{d})} }\!\!\right\}, \forall k \in {\cal K},$
\Until $\sum\limits_{k =1}^K \left|\ln w_k - \ln w_k'\right|\leq \rho_1,$
\vspace{0.3em}
\State $p_{{\text s},k}=f_k^2, \forall k \in {\cal K}.$
\label{algorithm2}
\end{algorithmic}
\end{algorithm}

Based on the above analysis, we can alternatively optimize ${\bm q}_{\text s}$ and ${\bm p}_{\text s}$ by iteratively applying the proposed DPCC and DPCD algorithms, and a suboptimal solution of problem (\ref{power_control}) can be obtained. Let JDPC denote the joint data power control algorithm which iteratively carries out DPCC and DPCD until convergence. For brevity, we omit the detailed description of the JDPC Algorithm.

\subsection{Convergence Analysis for the JDPC Algorithm}
Since we propose to solve the original problem (\ref{power_control}) using the JDPC algorithm, which operates in an iterative mechanism, it is necessary to characterize its convergence. In each iteration, the optimal ${\bm q}_{\text s}$ is first obtained by using the DPCC algorithm with fixed ${\bm p}_{\text s}$. Then, for determined ${\bm q}_{\text s}$, the DPCD algorithm outputs a suboptimal ${\bm p}_{\text s}$. As a result, the sum SE of D2D links increases in each iteration. Noting the fact that the sum SE of D2D links is always upper bounded, we can conclude that the JDPC algorithm converges to a suboptimal solution of problem (\ref{power_control}).
\subsection{Implementation and Complexity Analysis}
Based on the above analysis, within a coherence interval, each transmitter first transmits its pilot with the maximum power for channel estimation, and then transmits its signal using the power obtained by solving (\ref{power_control}). To successfully implement the proposed PSA and JDPC algorithms, BS requires the information of large-scale fading coefficients $u_n^{(\text{c})}, u_k^{(\text{d})}, v_{nk}^{(\text{c})}, v_{ik}^{(\text{d})}, \forall n \in {\cal N}, \forall i,k \in {\cal K}$. $u_n^{(\text{c})}, u_k^{(\text{d})}, \forall n \in {\cal N}$ can be directly estimated by BS, whereas $v_{nk}^{(\text{c})}, v_{ik}^{(\text{d})}, \forall n \in {\cal N}, \forall i,k \in {\cal K}$ need to be estimated at all D2D-Rxs and fed back to BS. After collecting these information, BS runs the pilot scheduling and power control algorithms, and then sends the results to all users. Since large-scale fading coefficients vary slowly, the proposed algorithms can be carried out at a coarser frame level granularity.

In the following, we analyze the computational complexity of the PSA and JDPC algorithms. Since the PSA algorithm has a complexity of ${\cal O}(K^2)$ per iteration and it is carried out $K$ times, the total complexity is ${\cal O}(K^3)$. As for the JDPC algorithm, assume that it is carried out $L_1$ times, and in each loop $L_2$ and $L_3$ iterations are required for the DPCC and DPCD algorithms to converge. The major complexity of the DPCC algorithm lies in computing  ${\bm F}{\bm q}_{\text s}$, which involves a complexity of ${\cal O}(N^2)$. Hence, the total complexity of the DPCC algorithm is ${\cal O}(L_2 K^2)$. Since the complexity of obtaining the Lagrange multipliers using the bisection method for accuracy $\rho$ is ${\cal O}(\log_2(1/\rho))$, the overall complexity of solving problem (\ref{power_control2}) using the DPCD algorithm is ${\cal O}(L_3 \log_2(1/\rho))$. Therefore, the JDPC algorithm has a total complexity of ${\cal O}(L_1 (L_2 K^2+ L_3 \log_2(1/\rho)))$. Simulation results show that $L_1$, $L_2$ and $L_3$ are small, so the proposed algorithms involve low complexity for efficient solutions.

\section{Simulation Results}
In this section, we present simulation results to evaluate the performance of the D2D underlaid massive MIMO system with PR. We consider a network in a $1000 \,\text{m}\times 1000 \,\text{m}$ square area and all transmitters are located uniformly in this cell. The distance between a D2D-Tx and its associated receiver is uniformly distributed in the range of $[0\,\text{m}, D_{\max}\,\text{m}]$. For brevity, we assume equal minimum SINR requirement for CUs, i.e., $\gamma_n = \gamma, \forall n\in {\cal N}$, and equal maximum power constraint for all transmitters, i.e., $Q_n=P_k=P, \forall n\in {\cal N}, k\in {\cal K}$. According to Theorem \ref{theorem4}, with a relatively low PR ratio, all D2D-Txs can transmit their pilot signals in the maximum power to increase the estimation accuracy. Therefore, in the following, we assume that $q_{{\text p},n}=p_{{\text p},k}=\tau P, \forall n\in {\cal N}, k\in {\cal K}$. Unless otherwise specified, the other system parameters are summarized in Table I. All simulation results are obtained by averaging over $10^4$ channel realizations, and each channel realization is obtained by generating a random user distribution as well as a random set of fading coefficients.
\subsection{Spectral Efficiency Versus the Corresponding Lower Bound}
\label{section6}
\begin{table}\centering
\label{Simu_Para}
\caption{Simulation Parameters}
\vspace{-0.8em}
\begin{tabular}{|l|l|}
\hline Maximum data transmit power $P$ & 17 dBm\\
\hline Additive noise power $N_0$ & -100 dBm\\
\hline Path loss exponent & 3.7\\
\hline Standard deviation of log-normal shadowing fading& 8 dB\\
\hline Accuracy $\rho$, $\rho_1$ & $10^{-3}$\\
\hline
\end{tabular}
\end{table}

\begin{figure}
  \centering
  \includegraphics[scale=0.45]{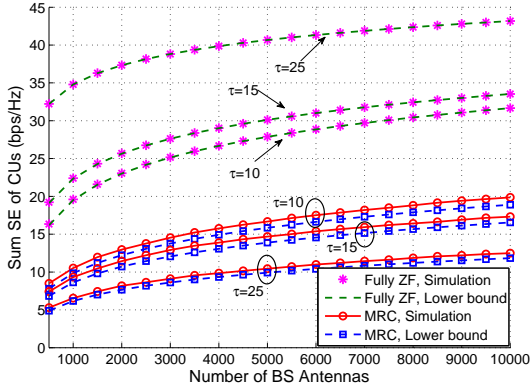}
  \caption{Simulated cellular sum SE and its lower bound versus the number of BS antennas with $q_{{\text s},n}=p_{{\text s},k}=P, \forall n \in {\cal N}, k \in {\cal K}$, $N=5$, $K=20$, $M=8$, $T=50$ and $D_{\max}=100$.}
  \label{Fig.1_ave_sum_rate_CU}
\end{figure}

In this subsection, we evaluate the tightness between the simulated SE and the corresponding lower bound of both cellular and D2D links.

First, we compare the simulated cellular sum SE with the corresponding lower bound under different values of pilot length and different receive filters at BS in Fig. \ref{Fig.1_ave_sum_rate_CU}. It can be seen that cellular sum SE increases with the number of BS antennas for all considered configurations, and fully ZF receivers (i.e., $(b_\text{c}, b_\text{d})=(N-1,\tau-N)$) greatly outperform MRC receivers (i.e., $(b_\text{c}, b_\text{d})=(0,0)$) in terms of cellular SE. As pilot length $\tau$ grows, cellular sum SE decreases for the MRC case, while increases for the fully ZF case. This has been explained in Remark \ref{remark1}. Moreover, Fig. \ref{Fig.1_ave_sum_rate_CU} also shows that the lower bound of cellular sum SE closely matches the simulation for the MRC case, and almost coincides with the simulation for the fully ZF case.
\begin{figure}
  \centering
  \includegraphics[scale=0.45]{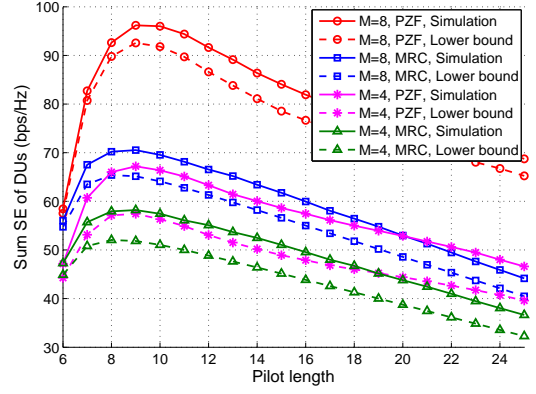}
  \caption{Simulated sum SE of D2D links and its lower bound versus pilot length with $q_{{\text s},n}=p_{{\text s},k}=P, \forall n \in {\cal N}, k \in {\cal K}$, $N=5$, $K=20$, $B=1024$, $T=50$ and $D_{\max}=100$.}
  \label{Fig.2_ave_sum_rate_DU}
\end{figure}

Next, in Fig. \ref{Fig.2_ave_sum_rate_DU}, we compare the simulated sum SE of D2D links with the corresponding lower bound under different numbers of D2D-Rx antennas and different receive filters. Note that when D2D-Rxs adopt PZF receivers for signal detection, as mentioned in Section \ref{Section_IV}, $(m_\text{c}, m_\text{d})$ should be in feasible set (\ref{feasible_m}). Therefore, in Fig. \ref{Fig.2_ave_sum_rate_DU}, we assume that $\left(m_\text{c}, m_\text{d}\right)=\left(1, \min \{ \tau -6, 1\}\right)$ when $M=4$, and $\left(m_\text{c}, m_\text{d}\right)=\left(1, \min \{ \tau -6, 2\}\right)$ when $M=8$. It can be seen from this figure that PZF receivers greatly outperform MRC receivers in terms of D2D sum SE. When $M=4$, the lower bound on the sum SE of DUs obtained by PZF receivers is approximate to the simulated SE obtained by MRC receivers. The gap between the simulated SE and its lower bound is small for all considered configurations, indicating that it is feasible to solve the data power control problem in Section \ref{section_Power_Control} based on lower bounds (\ref{SINR_lower_bound_CU}) and (\ref{lower_bound}). Fig. \ref{Fig.2_ave_sum_rate_DU} also shows that when $\tau \leq 9$, D2D sum SE increases with pilot length, while after that D2D sum SE decreases almost linearly with pilot length. This is because only a few orthogonal pilots are reused among D2D pairs for a small $\tau$. In this case, the channel estimation is significantly influenced by pilot contamination. Therefore, the SE of D2D links can be enhanced by increasing $\tau$. However, as $\tau$ becomes large enough, the channel estimation accuracy can be hardly improved by further enlarging $\tau$. Counterproductively, increasing pilot length reduces the number of symbols available for data transmission, and thereby decreases the SE of D2D links.

In the following simulation, we assume that BS applies fully ZF receivers for signal detection. As for D2D communication, since the number of D2D-Rx antennas may be smaller than the number of all transmitters in the cell, we assume that D2D-Rxs apply PZF receivers for signal detection. In addition,we set $B=1024$ and $M=8$.
\subsection{Performance of the PSA Algorithm}
\begin{figure}
  \centering
  \includegraphics[scale=0.45]{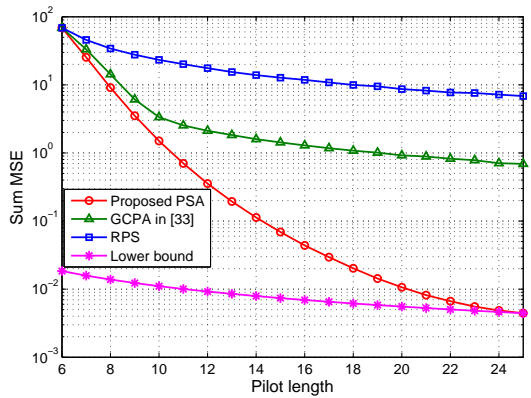}
  \caption{Performances of different pilot scheduling algorithms versus pilot length with $N=5$, $K=20$ and $D_{\max}=100$.}
  \label{Fig.3_Sum_MSE}
\end{figure}

\begin{figure}
  \centering
  \includegraphics[scale=0.45]{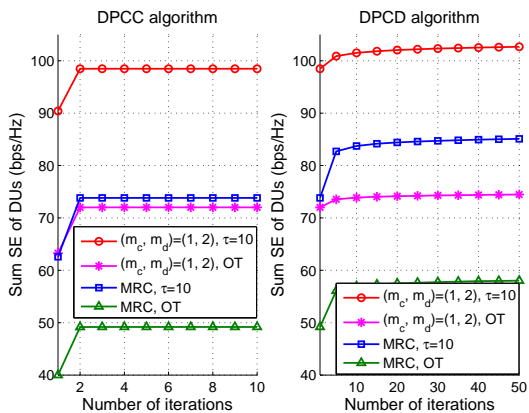}
  \caption{Convergence behaviors of the inner updates in the first outer iteration of the JDPC algorithm with $N=5$, $K=20$, $T=50$, $\gamma=5$ dB and $D_{\max}=100$.}
  \label{Fig.4_iteration_inner}
\end{figure}
In this subsection, we evaluate the performance of the proposed PSA algorithm. For comparison, in Fig. \ref{Fig.3_Sum_MSE}, we plot sum MSE (\ref{sum_MMSE}) versus pilot length for different pilot scheduling algorithms: the proposed PSA scheme, the GCPA algorithm proposed in \cite{zhu2015graph} and the random pilot scheduling (RPS). As a benchmark, we also depict the lower bound on the sum MSE, which is obtained when all users apply different orthogonal pilots for channel estimation. It can be observed from Fig. \ref{Fig.3_Sum_MSE} that the sum MSE decreases with pilot length for all algorithms, which is consistent with intuition. Moreover, the proposed PSA scheme approaches the lower bound quickly as $\tau$ increases, and outperforms the other two algorithms significantly in terms of sum MSE.

\subsection{Performance of the JDPC Algorithm}
In this subsection, we evaluate the performance of the proposed JDPC algorithm. Denote the orthogonal training scheme (i.e., $\tau = N+K$) by ` OT '.

\begin{figure}
  \centering
  \includegraphics[scale=0.45]{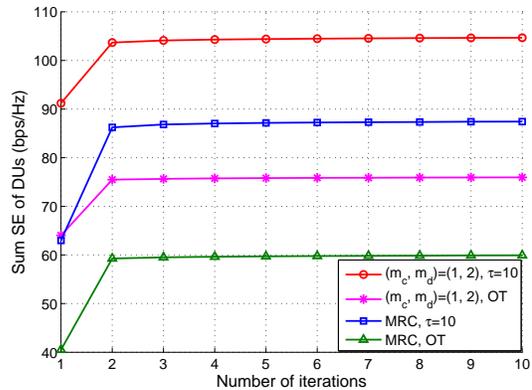}
  \caption{Convergence behaviors of the JDPC algorithm with $N=5$, $K=20$, $T=50$, $\gamma=5$ dB and $D_{\max}=100$.}
  \label{Fig.5_Sum_rate_DU_ite}
\end{figure}

Fig. \ref{Fig.4_iteration_inner} and Fig. \ref{Fig.5_Sum_rate_DU_ite} illustrate the convergence behaviors of the proposed power control algorithm under different configurations. Specifically, the left and right panels of Fig. \ref{Fig.4_iteration_inner} respectively correspond to the DPCC algorithm and the DPCD algorithm, while Fig. \ref{Fig.5_Sum_rate_DU_ite} corresponds to the JDPC algorithm. It can be seen from Fig. \ref{Fig.4_iteration_inner} that the sum SE of D2D links monotonically increases during the iterative procedure and converges rapidly for both DPCC and DPCD algorithms. By iteratively carrying out these two algorithms to optimize the data transmit power of CUs and DUs, the sum SE of D2D links can be effectively increased. Fig. \ref{Fig.5_Sum_rate_DU_ite} shows that the JDPC algorithm converges after only a few iterations (within 3 iterations for all considered configurations). This makes the proposed data power control algorithm suitable for practical applications.

\begin{figure}
  \centering
  \includegraphics[scale=0.45]{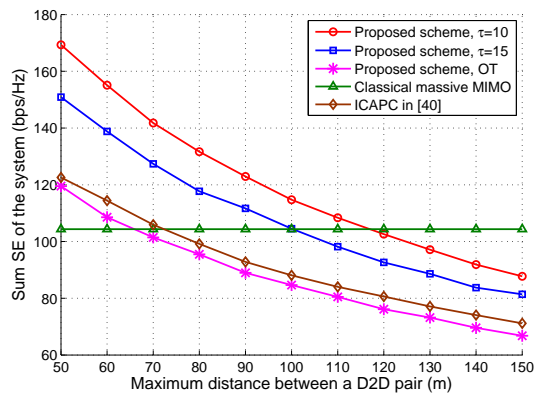}
  \caption{Sum SE of the system versus the maximum distance between a D2D pair with $(m_\text{c}, m_\text{d})=(1,2)$, $N=5$,  $K=20$, $T=50$ and $\gamma=5$ dB.}
  \label{Fig.6_sum_rate_DU_VS_distance}
\end{figure}

In Fig. \ref{Fig.6_sum_rate_DU_VS_distance}, the sum SE of the system versus the maximum distance between a D2D pair is depicted. We compare the proposed scheme with the iterative channel allocation and power control (ICAPC) algorithm proposed in \cite{xu2016channel}. Note that \cite{xu2016channel} considered a D2D underlaid cellular system with each transceiver equipped with one antenna. Therefore, to be fair, we set the same system configurations when simulating the ICAPC algorithm. We also include the sum SE curve obtained by using classical massive MIMO communication as a benchmark in Fig. \ref{Fig.6_sum_rate_DU_VS_distance}. To obtain this benchmark, instead of applying direct communication between D2D pairs, data signals of D2D-Txs are first forwarded to BS and then sent to D2D-Rxs. Let $R_n^{(\text{c})}$ denote the uplink SE of cellular link $n$, and $R_k^{(\text{d})}$ denote the SE of the transmission from D2D-Tx $k$ to D2D-Rx $k$ with BS working as the relay. Then, the benchmark can be obtained by maximizing $R=\sum \limits_{n=1}^N R_n^{(\text{c})}+\sum \limits_{k=1}^K R_k^{(\text{d})}$. When fully ZF receivers are applied at the BS, the maximum $R$ can be obtained by letting all transmitters using the maximum power for signal transmission. As expected, Fig. \ref{Fig.6_sum_rate_DU_VS_distance} shows that the sum SE of the system decreases with $D_{\max}$ when D2D communication is adopted. Since the ICAPC algorithm assumed perfect CSI and aimed to maximize the sum SE of DUs based on instantaneous CSI, the ICAPC algorithm outperforms the proposed scheme slightly in terms of system throughput when OT is adopted. However, with PR among DUs, the proposed scheme can obviously increase the system SE. Moreover, we can also conclude from Fig. \ref{Fig.6_sum_rate_DU_VS_distance} that due to the property of short-distance transmission, D2D communication can help improve the sum SE of a massive MIMO system significantly especially when $D_{\max}$ is small.

\begin{figure}
  \centering
  \includegraphics[scale=0.45]{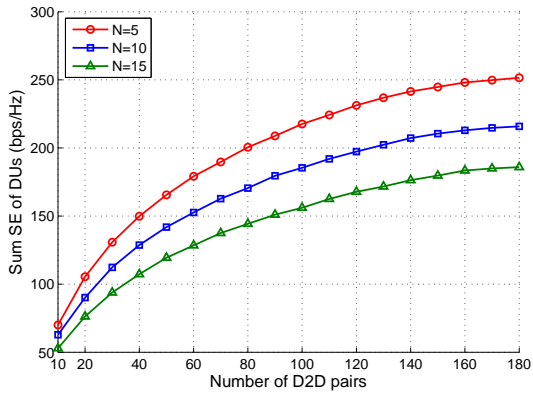}
  \caption{Sum SE of D2D links versus the number of D2D pairs under different values of coherence block length with $(m_\text{c}, m_\text{d})=(1,2)$, $\tau=10$, $T=50$, $\gamma=5$ dB and $D_{\max}=100$.}
  \label{Fig.7_sum_rate_DU_VS_K}
\end{figure}

\begin{figure}
  \centering
  \includegraphics[scale=0.45]{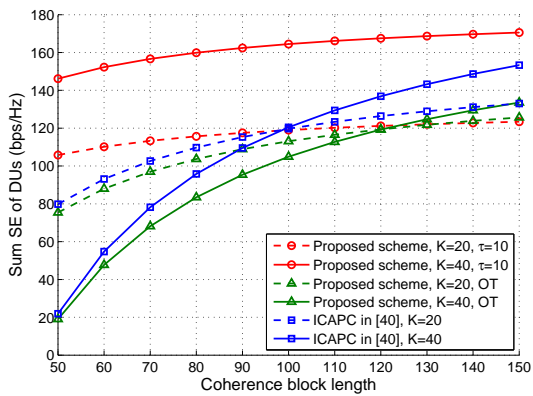}
  \caption{Sum SE of D2D links versus the coherence block length with $(m_\text{c}, m_\text{d})=(1,2)$, $N=5$, $\gamma=5$ dB and $D_{\max}=100$.}
  \label{Fig.8_sum_rate_DU_VS_T}
\end{figure}

Fig. \ref{Fig.7_sum_rate_DU_VS_K} depicts the sum SE of D2D links versus the number of D2D pairs under different values of $N$. It can be seen that the sum SE of DUs first increases with $K$ and then approaches a saturation point when $K>150$. This is because we set $\tau=10$, as $K$ grows, PR ratio increases, resulting in more pilot contamination for D2D channel estimations. Moreover, since a larger $N$ results in more cochannel interference and requires longer pilot overhead, as Fig. \ref{Fig.7_sum_rate_DU_VS_K} shows, the sum SE of DUs decreases with $N$.

\begin{figure}
  \centering
  \includegraphics[scale=0.45]{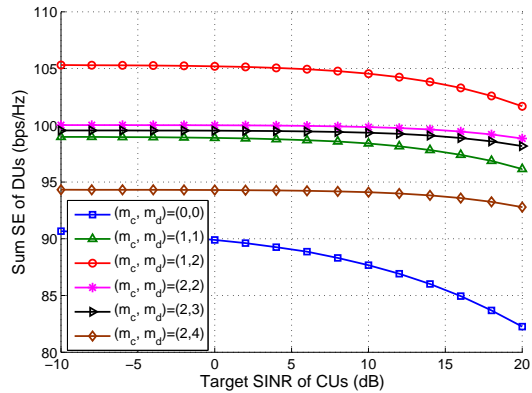}
  \caption{Sum SE of D2D links versus the target SINR of CUs under different $(m_\text{c}, m_\text{d})$ with $N=5$, $K=20$ , $\tau=10$, $T=50$ and $D_{\max}=100$.}
  \label{Fig.9_sum_rate_DU_VS_SINR}
\end{figure}
The impact of the coherence block length $T$ on the sum SE of DUs is investigated in Fig. \ref{Fig.8_sum_rate_DU_VS_T}. As expected, the sum SE of DUs increases with $T$ for all considered configurations. For both the proposed scheme with OT and the ICAPC algorithm, as $T$ grows, the sum SE of DUs with $K=40$ is first lower and then higher than that of the $K=20$ case. This is because with OT, in the small $T$ regime, the sum SE of DUs is mainly influenced by $\tau$, while in the high $T$ regime, it is mainly affected by $K$. Moreover, Fig. \ref{Fig.8_sum_rate_DU_VS_T} also shows that compared with OT, the sum SE of DUs can be greatly increased by PR especially when $T$ is small.

Fig. \ref{Fig.9_sum_rate_DU_VS_SINR} illustrates how the target SINR of CUs and PZF parameters $(m_\text{c}, m_\text{d})$ affect the sum SE of D2D links. Several observations can be made from this figure. First, the sum SE of D2D links decreases with $\gamma$, and the loss in D2D sum SE resulted from the increase of $\gamma$ reduces as $m_\text{c}$ grows from 0 to 2. Second, PZF receivers (i.e., $m_\text{c}+ m_\text{d}>0$) outperform the MRC receiver (i.e., $m_\text{c}= m_\text{d}=0$)  in terms of D2D sum SE for all different choices of $(m_\text{c}, m_\text{d})$. Moreover, for the considered configuration, the maximum sum SE of D2D links is obtained when $(m_\text{c}, m_\text{d})=(1,2)$. Further increasing the degrees of freedom for interference suppression results in decrease of D2D sum SE since less degrees of freedom are left for signal enhancement.

\section{Conclusions}
\label{section7}
In this paper, we consider a D2D underlaid massive MIMO system. Due to the fact that D2D pairs are usually located dispersively and conduct short-distance transmission in low power, letting several D2D pairs far from each other use the same pilot for channel estimation would be feasible and beneficial. Hence, we allow PR among DUs to reduce pilot overhead. Based on this setup, we first investigate the channel estimation under PR and derive a lower bound on the ergodic achievable rate of each link. To mitigate pilot contamination caused by PR, we develop a pilot scheduling algorithm under the criterion of minimizing the sum MSE of channel estimation of D2D links. In addition, we also maximize the sum rate of all D2D links based on the large-scale fading coefficients instead of the instantaneous CSI. An iterative power control algorithm is proposed to obtain a suboptimal solution. Simulation results show that the effect of pilot contamination can be decreased greatly by exploiting the proposed pilot scheduling algorithm, and the PR scheme can provide significant performance gains over the conventional orthogonal training scheme in terms of system SE.
\section*{Appendix A\\Proof of Theorem \ref{theorem2}}
By definition, $\left(\!{\bm \beta}_n^{(\text{c})}\!\right)^H {\hat {\bm h}}_n^{(\text{c})}$, $\left(\!{\bm \beta}_n^{(\text{c})}\!\right)^H {\hat {\bm h}}_a^{(\text{c})}$ and $\left(\!{\bm \beta}_n^{(\text{c})}\!\right)^H {\hat {\bm h}}_i^{(\text{d})}$ are respectively the projections of channel estimation vectors ${\hat {\bm h}}_n^{(\text{c})}$, ${\hat {\bm h}}_a^{(\text{c})}$ and ${\hat {\bm h}}_i^{(\text{d})}$ onto PZF receiver ${\bm \beta}_n^{(\text{c})}$. As mentioned in Section \ref{section3}, channel estimation of cellular links is only affected by additive noise. Hence, ${\hat {\bm h}_n^{(\text{c})}}$ is independent of ${\hat{\bm h}_a^{(\text{c})}}$ and ${\hat{\bm h}_i^{(\text{d})}}$, $\forall n \!\in\! {\cal N}, a \!\in\! {\cal N} \setminus n, i \!\in\! {\cal K}$. Then, from (\ref{SINR_CU}), we can find that the desired signal of CU $n$ $S_n^{(\text{c})}$ doesn't dependent on cochannel cellular and D2D interference $I_n^{(\text{c}\rightarrow \text{c})}$ and $I_n^{(\text{d}\rightarrow \text{c})}$. Using the convexity of $\log_2 \left(1+\frac{1}{x}\right)$ ($\forall x >0$) and applying the Jensen's inequality, we have
{\setlength\arraycolsep{2pt}
\begin{eqnarray}
&&{\mathbb E}\left( R_n^{(\text{c})}\right) \geq R_n^{(\text{c},{\text {lb}})}\nonumber\\
&&\triangleq \left(1-\frac{\tau}{T}\right)\log_2 \left(1+\left({\mathbb E}\left\{\frac{1}{\eta_n^{(\text{c})}}\right\}\right)^{-1}\right),\forall n \!\in \!{\cal N},\;\;
\label{jensen1}
\end{eqnarray}}
\!\!\!where $R_n^{(\text{c})}$ and $R_n^{(\text{c},{\text {lb}})}$ denote the instantaneous rate and a lower bound on the ergodic achievable rate of cellular link $n$, respectively. Based on (\ref{SINR_CU}) and the above analysis, we have
\begin{equation}
{\mathbb E}\left\{\frac{1}{\eta_n^{(\text{c})}}\right\}={\mathbb E}\left\{\frac{1}{S_n^{(\text{c})}}\right\} \left({\mathbb E}\left\{ I_n^{(\text{c}\rightarrow \text{c})}\right\}+ {\mathbb E}\left\{I_n^{(\text{d}\rightarrow \text{c})}\right\}+ \alpha^{(\text{c})}\right).
\label{expectation_CU}
\end{equation}

In order to obtain an explicit expression of (\ref{expectation_CU}), we need to calculate ${\mathbb E}\left\{\frac{1}{S_n^{(\text{c})}}\right\}$, ${\mathbb E}\left\{ I_n^{(\text{c}\rightarrow \text{c})}\right\}$ and ${\mathbb E}\left\{I_n^{(\text{d}\rightarrow \text{c})}\right\}$. For convenience, denote $\bm G = [\bm G_1, \bm G_2] \in {\mathbb C}^{B\times B}$, where the columns of $\bm G$ form an orthogonal basis of the $B$ dimensional space. Specifically, the columns of $\bm G_2\in {\mathbb C}^{B\times (b_\text{c} + b_\text{d})}$ form an orthogonal basis of the subspace spanned by channel estimation vectors of cancelled interferers, and each column of $\bm G_1\in {\mathbb C}^{B\times (B - b_\text{c} - b_\text{d})}$ is orthogonal to $\bm G_2$. Then, the unit norm PZF receiver ${\bm \beta}_n^{(\text{c})}$ can be chosen as
\begin{equation}
{\bm \beta}_n^{(\text{c})}=\frac{\bm G_1\bm G_1^H {\hat {\bm h}}_n^{(\text{c})}}{\left\|\bm G_1\bm G_1^H {\hat {\bm h}}_n^{(\text{c})} \right\|_2}.
\label{PZF_CU}
\end{equation}
By expressing ${\hat {\bm h}}_n^{(\text{c})}$ as the sum of projections of ${\hat {\bm h}}_n^{(\text{c})}$ onto $\bm G_1$ and $\bm G_2$, we have
{\setlength\arraycolsep{2pt}
\begin{eqnarray}
\left({\bm \beta}_n^{(\text{c})}\right)^H {\hat {\bm h}}_n^{(\text{c})} &=& \left({\bm \beta}_n^{(\text{c})}\right)^H \left(\bm G_1\bm G_1^H {\hat {\bm h}}_n^{(\text{c})} +\bm G_2\bm G_2^H {\hat {\bm h}}_n^{(\text{c})} \right) \nonumber\\
&=& \left\|\bm G_1 \bm G_1^H {\hat {\bm h}}_n^{(\text{c})}\right\|_2\nonumber\\
&=& \left\|\bm G_1^H {\hat {\bm h}}_n^{(\text{c})}\right\|_2,
\label{change1}
\end{eqnarray}}
\!\!\!which indicates that $\left({\bm \beta}_n^{(\text{c})}\right)^H {\hat {\bm h}}_n^{(\text{c})}$ equals to the norm of the projection of ${\hat {\bm h}}_n^{(\text{c})}$ on $\bm G_1$. Obviously, ${\hat {\bm h}}_n^{(\text{c})}$ is independent of each column of $\bm G_1$. Hence, each element of $\bm G_1^H {\hat {\bm h}}_n^{(\text{c})}$ follows i.i.d. ${\cal {CN}}(0,\delta_n^{(\text{c})})$. According to \cite[Section~2.1.6]{tulino2004random}, $\frac{1}{\delta_n^{(\text{c})}} \left\|\bm G_1^H {\hat {\bm h}}_n^{(\text{c})}\right\|_2^2 \sim {\cal W}_1 (B- b_\text{c} - b_\text{d},1)$ is a central complex Wishart random variable with $B- b_\text{c} - b_\text{d}$ degrees of freedom. Then, we have
\begin{equation}
{\mathbb E} \left\{ \frac{1}{S_n^{(\text{c})}}\right\}=\frac{1}{q_{{\text s},n} u_n^{(\text{c})} (B- b_\text{c} - b_\text{d} -1)\delta_n^{(\text{c})}}.
\label{S_CU}
\end{equation}

Since ${\bm \beta}_n^{(\text{c})}$ is an unit norm vector and is independent of ${\hat {\bm h}}_a^{(\text{c})}, \forall a \in {\cal C}_n^{(\text{c})} \setminus n$, $\left({\bm \beta}_n^{(\text{c})}\right)^H {\hat {\bm h}_a^{(\text{c})}}$ is the linear combination of complex Gaussian random variables and follows ${\cal {CN}}(0,\delta_a^{(\text{c})})$. Hence, $\left|\left({\bm \beta}_n^{(\text{c})}\right)^H {\hat {\bm h}_a^{(\text{c})}}\right|^2$ follows exponential distribution, i.e., $\frac{1}{\delta_a^{(\text{c})}}\left|\left({\bm \beta}_n^{(\text{c})}\right)^H {\hat {\bm h}_a^{(\text{c})}}\right|^2 \sim {\text {Exp}}(1), \forall a \in {\cal C}_n^{(\text{c})} \setminus n$. Similarly, $\frac{1}{\delta_i^{(\text{d})}}\left|\left({\bm \beta}_n^{(\text{c})}\right)^H {\hat {\bm h}_i^{(\text{d})}}\right|^2 \sim {\text {Exp}}(1), \forall i \in {\cal D}^{(\text{c})}$. Then, we have
{\setlength\arraycolsep{2pt}
\begin{eqnarray}
{\mathbb E} \left\{ I_n^{(\text{c}\rightarrow \text{c})}\right\}&=&\sum\limits_{a \in {\cal C}_n^{(\text{c})}\setminus n} q_{{\text s},a} u_a^{(\text{c})} {\mathbb E} \left\{ \left|\left({\bm \beta}_n^{(\text{c})}\right)^H {\hat {\bm h}_a^{(\text{c})}}\right|^2\right\}\nonumber\\
&=& \sum\limits_{a \in {\cal C}_n^{(\text{c})}\setminus n} q_{{\text s},a} u_a^{(\text{c})} \delta_a^{(\text{c})}, \nonumber\\
{\mathbb E} \left\{I_n^{(\text{d}\rightarrow \text{c})}\right\}&=& \sum\limits_{i\in {\cal D}^{(\text{c})}} p_{{\text s},i} u_i^{(\text{d})}  {\mathbb E} \left\{\left|\left({\bm \beta}_n^{(\text{c})}\right)^H {\hat {\bm h}_i^{(\text{d})}}\right|^2\right\}\nonumber\\
&=& \sum\limits_{i\in {\cal D}^{(\text{c})}} p_{{\text s},i} u_i^{(\text{d})} \delta_i^{(\text{d})}.
\label{I_CU}
\end{eqnarray}}
\!\!Substituting (\ref{S_CU}) and (\ref{I_CU}) into (\ref{jensen1}) yields lower bound (\ref{lower_bound_CU}).
\section*{Appendix B\\Proof of Theorem \ref{theorem3}}
Unlike the channel estimation of cellular links which is only affected by additive noise, pilot contamination exists when estimating D2D channels due to PR. As a result, the estimation of ${\bm g}_{kk}^{(\text{d})}$ is not only affected by noise, but also influenced by pilot contamination caused by DUs $i \in {\cal X}_k \setminus k$. From (\ref{channel_estimation}), we can obtain the following relationship
\begin{equation}
{\hat {\bm g}_{ik}^{(\text{d})}} = \sqrt {\frac{\mu_{ik}^{(\text{d})}}{\mu_{kk}^{(\text{d})}}} {\hat {\bm g}_{kk}^{(\text{d})}}, \forall i \in {\cal X}_k \setminus k.
\label{g_rela}
\end{equation}
Denote
{\setlength\arraycolsep{2pt}
\begin{eqnarray}
I_k^{(\text{d}\rightarrow \text{d})} &=& {\hat I}_k^{(\text{d}\rightarrow \text{d})} + {\tilde I}_k^{(\text{d}\rightarrow \text{d})}, \nonumber\\
{\hat I}_k^{(\text{d}\rightarrow \text{d})} &=& \sum\limits_{i \in {\cal D}_k^{(\text{d})}\setminus {\cal X}_k} p_{{\text s},i} v_{ik}^{(\text{d})} \left|\left({\bm \beta}_k^{(\text{d})}\right)^H {\hat {\bm g}_{ik}^{(\text{d})}}\right|^2, \nonumber\\
{\tilde I}_k^{(\text{d}\rightarrow \text{d})} &=& \sum\limits_{i \in {\cal X}_k \setminus k} p_{{\text s},i} v_{ik}^{(\text{d})} \left|\left({\bm \beta}_k^{(\text{d})}\right)^H {\hat {\bm g}_{ik}^{(\text{d})}}\right|^2.
\label{I_seperate}
\end{eqnarray}}
\!\!\!Then, from (\ref{g_rela}), we have
\begin{equation}
\frac{{\tilde I}_k^{(\text{d}\rightarrow \text{d})}}{S_k^{(\text{d})}}= \frac{\sum\limits_{i \in {\cal X}_k \setminus k} p_{{\text s},i} v_{ik}^{(\text{d})} \mu_{ik}^{(\text{d})}}{p_{{\text s},k} v_{kk}^{(\text{d})} \mu_{kk}^{(\text{d})}}.
\label{I_S}
\end{equation}

Analogous to the proof process in Appendix A, we can easily verify that $S_k^{(\text{d})}$ is independent of ${\hat I}_k^{(\text{d}\rightarrow \text{d})}$ and $I_k^{(\text{c}\rightarrow \text{d})}$. Hence, the relationship between the ergodic achievable rate of D2D link $k$ and its lower bound can be expressed as
{\setlength\arraycolsep{2pt}
\begin{eqnarray}
&&{\mathbb E}\left( R_k^{(\text{d})}\right) \!\geq \! R_k^{(\text{d},{\text {lb}})} \nonumber\\
&&\triangleq \left(1-\frac{\tau}{T}\right)\log_2 \left(1\!+\!\left({\mathbb E}\left\{\frac{1}{\eta_k^{(\text{d})}}\right\}\right)^{-1}\right),\forall k \!\in \!{\cal K},
\label{jensen}
\end{eqnarray}}
\!\!\!where
{\setlength\arraycolsep{2pt}
\begin{eqnarray}
{\mathbb E}\left\{\frac{1}{\eta_k^{(\text{d})}}\right\}&=&{\mathbb E}\left\{\!\frac{1}{S_k^{(\text{d})}}\!\right\}\! \left({\mathbb E}\left\{ {\hat I}_k^{(\text{d}\rightarrow \text{d})}\right\}+ {\mathbb E}\left\{I_k^{(\text{c}\rightarrow \text{d})}\right\}+ \alpha_k^{(\text{d})}\right)\nonumber\\
&+&\frac{{\tilde I}_k^{(\text{d}\rightarrow \text{d})}}{S_k^{(\text{d})}}.
\label{expectation_DU}
\end{eqnarray}}
\!\!\!Besides, we also have
{\setlength\arraycolsep{2pt}
\begin{eqnarray}
&&{\mathbb E} \left\{ \frac{1}{S_k^{(\text{d})}}\right\}=\frac{1}{p_{{\text s},k} v_{kk}^{(\text{d})} (M- m_\text{c} - m_\text{d} -1)\mu_{kk}^{(\text{d})}}, \nonumber\\
&&{\mathbb E} \left\{I_k^{(\text{c}\rightarrow \text{d})}\right\}= \sum\limits_{n\in {\cal C}_k^{(\text{d})}} q_{{\text s},n} v_{nk}^{(\text{c})} \mu_{nk}^{(\text{c})}, \nonumber\\
&&{\mathbb E} \left\{{\hat I}_k^{(\text{d}\rightarrow \text{d})}\right\}= \sum\limits_{i \in {\cal D}_k^{(\text{d})}\setminus {\cal X}_k}  p_{{\text s},i} v_{ik}^{(\text{d})} \mu_{ik}^{(\text{d})}.
\label{S_DU}
\end{eqnarray}}
\!\!\!Substituting (\ref{I_S}) and (\ref{S_DU}) into (\ref{jensen}) yields lower bound (\ref{lower_bound}).
\section*{Appendix C\\Proof of Theorem \ref{theorem4}}
When the optimal pilot scheduling scheme has been determined to allocate pilots $\{{\bm \omega}_{N+1},\cdots,{\bm \omega}_{\tau}\}$ to DUs, problem (\ref{MMSE}) can be equivalently written as
{\setlength\arraycolsep{2pt}
\begin{subequations}
\begin{align}
\mathop {\max }\limits_{{\bm p}_{\text p}} \quad& \sum \limits_{k=1}^K \frac{U_k({\bm p}_{\text p})}{V_k({\bm p}_{\text p})} = \sum \limits_{k=1}^K \frac{p_{{\text p},k} v_{kk}^{(\text{d})}}{ \sum\limits_{i \in {\cal X}_k} p_{{\text p},i} v_{ik}^{(\text{d})}+N_0 }  \label{MMSE2_a}\\
\text{s.t.} \quad\; & 0 \leq p_{{\text p},k} \leq \tau P_k, \forall k \in {\cal K}. \label{MMSE2_b}
\end{align}
\label{MMSE2}
\end{subequations}}
\!\!\!\!\!Obviously, problem (\ref{MMSE2}) is a non-convex sum-of-ratios optimization, which aims to maximize the summation of fractional functions. The parametric algorithm is often adopted to solve this kind of problem if the numerator of each summation term is concave and the denominator of each summation term is convex\cite{jong2012efficient}, \cite{xu2015energy}. Since $U_k({\bm p}_{\text p})$ and $V_k({\bm p}_{\text p})$ are both affine w.r.t. ${\bm p}_{\text p}$ for any $k$, we can optimally solve problem (\ref{MMSE2}) using the parametric algorithm. According to \cite{jong2012efficient}, (\ref{MMSE2}) can be equivalently transformed to the following problem
{\setlength\arraycolsep{2pt}
\begin{subequations}
\begin{align}
\mathop {\max }\limits_{{\bm p}_{\text p}, {\bm \xi}} \quad& \sum \limits_{k=1}^K \xi_k  \label{MMSE3_a}\\
\text{s.t.} \quad\; & \frac{U_k({\bm p}_{\text p})}{V_k({\bm p}_{\text p})}\geq \xi_k, \forall k \in {\cal K},\label{MMSE3_b}\\
& 0 \leq p_{{\text p},k} \leq \tau P_k, \forall k \in {\cal K}, \label{MMSE3_c}
\end{align}
\label{MMSE3}
\end{subequations}}
\!\!\!\!where ${\bm \xi}\!=\!(\xi_1, \cdots, \xi_K)^T$. In order to solve the problem in (\ref{MMSE3}), we resort to alternating optimization by using the following Lemma. Applying \cite[Lemma 2.1]{jong2012efficient}, we obtain

\begin{lemma}
If $({\bm p}_{\text p}^{\text {opt}},{\bm \xi}^{\text {opt}})$ is the optimal solution of the above maximization problem in (\ref{MMSE3}), then there exists ${\bm \kappa}^{\text {opt}}=(\kappa_1^{\text {opt}}, \cdots, \kappa_K^{\text {opt}})^T$ such that ${\bm p}_{\text p}^{\text {opt}}$ is the optimal solution to the following problem
{\setlength\arraycolsep{2pt}
\begin{subequations}
\begin{align}
\mathop {\max }\limits_{{\bm p}_{\text p}} \quad& \sum \limits_{k=1}^K \kappa_k^{\text {opt}} \left( U_k({\bm p}_{\text p}) - \xi_k^{\text {opt}} V_k({\bm p}_{\text p}) \right)  \label{MMSE4_a}\\
\text{s.t.} \quad\; & 0 \leq p_{{\text p},k} \leq \tau P_k, \forall k \in {\cal K}. \label{MMSE4_b}
\end{align}
\label{MMSE4}
\end{subequations}}
\!\!\!\!Meanwhile, the optimal values of $({\bm \kappa}^{\text {opt}}, {\bm \xi}^{\text {opt}})$ should satisfy
{\setlength\arraycolsep{2pt}
\begin{eqnarray}
&& \kappa_k^{\text {opt}} = \frac{1}{V_k({\bm p}_{\text p}^{\text {opt}})}, \forall k \in {\cal K},\nonumber\\
&& U_k({\bm p}_{\text p}^{\text {opt}})-\xi_k^{\text {opt}} V_k({\bm p}_{\text p}^{\text {opt}}) = 0, \forall k \in {\cal K}.
\label{MMSE5}
\end{eqnarray}}
\label{lemma1}
\end{lemma}

From Lemma \ref{lemma1}, we can obtain $({\bm p}_{\text p}^{\text {opt}},{\bm \xi}^{\text {opt}},{\bm \kappa}^{\text {opt}})$ by iteratively carrying out the following two steps until convergence: 1) update $({\bm \xi},{\bm \kappa})$ based on (\ref{MMSE5}) for given ${\bm p}_{\text p}$; 2) update ${\bm p}_{\text p}$ for given $({\bm \xi},{\bm \kappa})$ by solving (\ref{MMSE4}). In the last iteration of the parametric algorithm, assume that $({\bm \xi}^{\text {opt}},{\bm \kappa}^{\text {opt}})$ has been obtained. Then, we find the final ${\bm p}_{\text p}^{\text {opt}}$ by solving (\ref{MMSE4}), which can be further reformulated as
{\setlength\arraycolsep{2pt}
\begin{subequations}
\begin{align}
\mathop {\max }\limits_{{\bm p}_{\text p}} \quad& \!\!\sum \limits_{k=1}^K \!\left( \!p_{{\text p},k} \!\left( \!\kappa_k^{\text {opt}} v_{kk}^{(\text{d})} \!-\! \sum\limits_{i \in {\cal X}_k} \!\kappa_i^{\text {opt}} \xi_i^{\text {opt}} v_{ki}^{(\text{d})} \!\right) \!-\! \kappa_k^{\text {opt}} \xi_k^{\text {opt}} N_0\!\right)  \label{MMSE7_a}\\
\text{s.t.} \quad\; & 0 \leq p_{{\text p},k} \leq \tau P_k, \forall k \in {\cal K}. \label{MMSE7_b}
\end{align}
\label{MMSE7}
\end{subequations}}
\!\!\!\!\!\!\!Since the above problem in (\ref{MMSE7}) is linear w.r.t. ${\bm p}_{\text p}$, it is directly known that $p_{{\text p},k}^{\text {opt}} \!=\! \tau P_k$ if $\kappa_k^{\text {opt}} v_{kk}^{(\text{d})} \!\geq\! \sum\limits_{i \in {\cal X}_k} \kappa_i^{\text {opt}} \xi_i^{\text {opt}} v_{ki}^{(\text{d})}$; Otherwise, $p_{{\text p},k}^{\text {opt}} \!=\! 0$. Notice that we have to guarantee $p_{{\text p},k}^{\text {opt}} > 0$ to estimate the channel vector of D2D link $k$. When $p_{{\text p},k}^{\text {opt}} = 0$, we can always change the pilot assigned to D2D pair $k$ by increasing $\tau$ so that $\kappa_k^{\text {opt}} v_{kk}^{(\text{d})} \geq \sum\limits_{i \in {\cal X}_k} \kappa_i^{\text {opt}} \xi_i^{\text {opt}} v_{ki}^{(\text{d})}$. The worst case is that $\tau=N+K$ and all D2D pairs use different orthogonal pilots for channel estimation. In this case, pilot contamination vanishes and each D2D-Tx would transmit its pilot with the maximum power to increase the channel estimation accuracy. Therefore, Theorem \ref{theorem4} is proven.
\section*{Appendix D\\Convergence Proof of of the DPCC Algorithm}
As discussed in \cite[Theorem 2]{yates1995framework}, the iterative process in (\ref{iteration}) converges to the optimal solution for any initial power vector ${\bm q}_{\text s} \succeq {\bm 0}$ if function ${\bm \Lambda} ({\bm q}_{\text s})$ is standard. Moreover, the sufficient condition for ${\bm \Lambda} ({\bm q}_{\text s})$ to be standard is that ${\bm \Delta}({\bm q}_{\text s})$ is standard. Therefore, we only need to prove that ${\bm \Delta}({\bm q}_{\text s})$ is standard. Since the elements in ${\bm F}$ are nonnegative, ${\bm q}_{\text s} \succeq {\bm 0}$ and ${\bm \theta} \succ {\bm 0}$, we have

$\bullet$ \emph{Positivity:} ${\bm \Delta}({\bm q}_{\text s})={\bm F}{\bm q}_{\text s} + {\bm \theta} \succ {\bm 0};$

$\bullet$ \emph{Monotonicity:} If ${\bm q}_{\text s} \succeq {\bm q}_{\text s}^{'}$, then ${\bm \Delta}({\bm q}_{\text s})-{\bm \Delta}({\bm q}_{\text s}^{'})={\bm F}({\bm q}_{\text s}-{\bm q}_{\text s}^{'})\succeq {\bm 0}$, i.e. ${\bm \Delta}({\bm q}_{\text s}) \succeq {\bm \Delta}({\bm q}_{\text s}^{'})$;

$\bullet$ \emph{Scalability:} $\forall \varsigma>1$, $\varsigma {\bm \Delta}({\bm q}_{\text s}) = \varsigma{\bm F}{\bm q}_{\text s} + \varsigma{\bm \theta}\succ \varsigma{\bm F}{\bm q}_{\text s} + {\bm \theta}= {\bm \Delta}(\varsigma{\bm q}_{\text s})$.

As a result, ${\bm \Lambda} ({\bm q}_{\text s})$ is standard and the iterative process in (\ref{iteration}) converges to the optimal solution for any initial power vector ${\bm q}_{\text s} \succeq {\bm 0}$.

\section*{Appendix E\\Solving (\ref{power_control2}) via WMMSE Algorithm}
For notational convenience, we denote $f_k= \sqrt{p_{{\text s},k}}, \forall k \in {\cal K}$. Then, (\ref{power_control2}) can be equivalently transformed into the following weighted sum-MSE minimization problem
{\setlength\arraycolsep{2pt}
\begin{subequations}
\begin{align}
\mathop {\min }\limits_{{\bm w},{\bm \nu},{\bm f}} \quad& \sum\limits_{k = 1}^K \left( w_k e_k - \ln w_k \right)  \label{equivalent_problem2_a}\\
\text{s.t.} \quad\; & \sum\limits_{k=1}^K f_k^2 \varphi_k^{(\text{d})} \leq \zeta,\label{equivalent_problem2_b}\\
& 0 \leq f_k \leq \sqrt{P_k}, \forall k \in {\cal K}, \label{equivalent_problem2_c}
\end{align}
\label{equivalent_problem2}
\end{subequations}}
\!\!\!\!\!where $w_k$ is a positive weight variable, and $e_k$ is the mean-square estimation error
\begin{equation}
e_k = \left(\nu_k f_k \sqrt{\phi_k^{(\text{d})}} - 1\right)^2 +\nu_k^2 \left(\sum\limits_{i =1}^K  f_i^2 \psi_{ik}^{(\text{d})} +  \sigma_k^{(\text{d})}\right).
\label{estimation_error}
\end{equation}

To prove the equivalence between (\ref{power_control2}) and (\ref{equivalent_problem2}), we derive the optimal ${\bm \nu}$ and ${\bm w}$ by checking the first-order optimality condition of problem (\ref{equivalent_problem2})
{\setlength\arraycolsep{2pt}
\begin{eqnarray}
\nu_k^\text {opt} &=& \frac{f_k \sqrt{\phi_k^{(\text{d})}}}{f_k^2 \phi_k^{(\text{d})}+\sum\limits_{i=1}^K f_i^2 \psi_{ik}^{(\text{d})} + \sigma_k^{(\text{d})}}, \forall k \in {\cal K},\nonumber\\
w_k^\text {opt} &=& \frac{1}{e_k}, \forall k \in {\cal K}.
\label{optimal_uw}
\end{eqnarray}}
Plugging (\ref{optimal_uw}) in (\ref{estimation_error}) and simplifying (\ref{equivalent_problem2}), we have
{\setlength\arraycolsep{2pt}
\begin{subequations}
\begin{align}
\mathop {\max }\limits_{\bm f} \quad& \sum \limits_{k=1}^K \log_2 \left( 1+\frac{f_k^2 \phi_k^{(\text{d})}}{\sum \limits_{i=1}^K f_i^2 \psi_{ik}^{(\text{d})} + \sigma_k^{(\text{d})}}\right)  \label{equivalent_problem3_a}\\
\text{s.t.} \quad\; & \sum\limits_{k=1}^K f_k^2 \varphi_k^{(\text{d})} \leq \zeta,\label{equivalent_problem3_b}\\
& 0 \leq f_k \leq \sqrt{P_k}, \forall k \in {\cal K}, \label{equivalent_problem3_c}
\end{align}
\label{equivalent_problem3}
\end{subequations}}
\!\!\!\!which is equivalent to (\ref{power_control2}). The equivalence implies that a suboptimal solution of problem (\ref{power_control2}) can be obtained by solving (\ref{equivalent_problem2}), which is easier to handle since the objective function is convex w.r.t. each variable when the other variables are fixed. The optimal ${\bm w}$ (or ${\bm \nu}$) can be obtained based on (\ref{optimal_uw}) when ${\bm \nu}$ and ${\bm f}$ (${\bm w}$ and ${\bm f}$) are fixed. For given ${\bm w}$ and ${\bm \nu}$, to get the optimal ${\bm f}$, we attach Lagrange multiplier $\lambda$ to the first constraint of (\ref{equivalent_problem2}) and obtain the Lagrange function as follows
\begin{equation}
{\cal L}({\bm f},\lambda) \triangleq  \sum\limits_{k = 1}^K \left(w_k e_k - \ln w_k\right)+ \lambda \left(\sum\limits_{k=1}^K f_k^2 \varphi_k^{(\text{d})} - \zeta\right).
\label{Lagrange_function}
\end{equation}
From \cite{boyd2004convex}, the first-order optimality condition of ${\cal L}$ w.r.t. $f_k$ yields
\begin{equation}
f_k(\lambda)= \min  \left\{{\sqrt {P_k}}, \frac{w_k \nu_k \sqrt{\phi_k^{(\text{d})}}}{w_k \nu_k^2 \phi_k^{(\text{d})} + \sum\limits_{i=1}^K w_i \nu_i^2 \psi_{ki}^{(\text{d})}+ \lambda \varphi_k^{(\text{d})} }\right\}.
\label{optimal_power}
\end{equation}
According to the complementary slackness condition, if $\lambda_{\text {opt}}=0$ yields $\sum\limits_{k=1}^K f_k^2 \varphi_k^{(\text{d})} < \zeta$, then, $f_k^\text {opt}=f_k(0)$. Otherwise, we have $\lambda_{\text {opt}}>0$ and $\sum\limits_{k=1}^K f_k^2 \varphi_k^{(\text{d})} = \zeta$. Since for any $k \in {\cal K}$, $f_k(\lambda)$ strictly decreases with $\lambda$, we can obtain $\lambda_{\text {opt}}$ using the bisection method. Then, substituting $\lambda_{\text {opt}}$ into (\ref{optimal_power}), we get $f_k^{\text {opt}}$.

\bibliographystyle{IEEEtran}
\bibliography{IEEEabrv,MMM}

\end{document}